\documentclass[sigconf,anonymous=false,review=false]{acmart}

\usepackage{url}  %
\usepackage{graphicx}  %
\usepackage{booktabs} %
\usepackage{hyperref}
\usepackage{tikz}
\usepackage{pgfplots}
\pgfplotsset{compat=1.13}
\usepgfplotslibrary{colormaps}
\usetikzlibrary{pgfplots.colormaps}
\usetikzlibrary{pgfplots.dateplot}
\usetikzlibrary{patterns}
\usepackage{filecontents}
\usepackage{graphicx}
\usepackage{subcaption}
\usepackage{color, colortbl}
\usepackage{arydshln}
\usepackage{float}
\usepackage[utf8]{inputenc}
\usepackage{arabtex}
\usepackage{utf8}
\usepackage{units}
\usepackage{enumitem}
\usepackage{balance}
\usepackage{microtype} %
\usepackage[export]{adjustbox}

\setcode{utf8}

\definecolor{Gray}{gray}{0.9}

\newcommand{\afblock}[1]{\noindent{\textbf{#1 }}}

\newcommand\encircle[1]{%
	\tikz[
		baseline={([yshift=-8pt]current bounding box.north)}
	]
		\node (X) [draw, shape=circle, inner sep=0, fill=black, text=white,scale=0.7] {\strut #1};
	\hspace{-4pt}
}

\newcommand{\ie}{i.e.,\,}
\newcommand{\Ie}{I.e.,\,}
\newcommand{\eg}{e.g.,\,}
\newcommand{\Eg}{E.g.,\,}
\newcommand{\etal}{et~al\@ifnextchar.{}{.\@}}
\newcommand{\etc}{etc\@ifnextchar.{}{.\@}}

\copyrightyear{2019}
\acmYear{2019}
\setcopyright{iw3c2w3}
\acmConference[WWW '19 Companion]{Companion Proceedings of the 2019 World Wide Web Conference}{May 13--17, 2019}{San Francisco, CA, USA}
\acmBooktitle{Companion Proceedings of the 2019 World Wide Web Conference (WWW '19 Companion), May 13--17, 2019, San Francisco, CA, USA}
\acmPrice{}
\acmDOI{10.1145/3308560.3316537}
\acmISBN{978-1-4503-6675-5/19/05}

\begin{document}
\title{Hashtag Usage in a Geographically-Local Microblogging App}

\renewcommand{\shortauthors}{J.H. Reelfs et al.}

\author{Helge Reelfs, Timon Mohaupt, Oliver Hohlfeld}
\affiliation{\institution{RWTH Aachen University}}

\author{Niklas Henckell}
\affiliation{\institution{The Jodel Venture GmbH}}

\begin{abstract}
This paper studies for the first time the usage and propagation of hashtags in a new and fundamentally different type of social media that is {\em i)} without profiles and {\em ii)} location-based to only show nearby posted content.
Our study is based on analyzing the mobile-only Jodel microblogging app, which has an established user base in several European countries and Saudi Arabia.
All posts are user to user anonymous (\ie no displayed user handles) and are only displayed in the proximity of the user's location (up to 20\,km).
It thereby forms local communities and opens the question of how information propagates within and between these communities.
We tackle this question by applying established metrics for Twitter hashtags to a ground-truth data set of Jodel posts within Germany that spans three years.
We find the usage of hashtags in Jodel to differ from Twitter; despite embracing local communication in its design, Jodel hashtags are mostly used country-wide.
\end{abstract}

\maketitle



\newcommand{\numberNumCharactersPerJodel}[0]{$250$}

\section{Introduction}

Social media has become a popular and ubiquitous tool for consuming and sharing digital content (\eg textual or multimedia).
This sharing leads to information propagation and spreading across users and even across different networks~\cite{zannettou2017web}.
Understanding this propagation has thus motivated research studies to investigate the dynamics of information adoption, spreading, and (complex) contagion of information~\cite{spatiotemp, epidemics_1,epidemics_2,epidemics_3, butterflies, local_variation, topics, facebook}, \eg in the form of memes.
A widely studied platform in this regard is the microblogging service Twitter that enables users to reach a global audience and for which sampled post data is available via APIs.
Analyzing the post contents' (\eg included memes) is, however, a very challenging application of natural language processing.
Since users often self-classify their posts by adding hashtags to ease retrieval, analyzing hashtags is a promising proxy measure for analyzing memes or post contents.
This has resulted in metrics to analyze hashtags and thereby valuable insights into their spreading behavior~\cite{spatiotemp}.

New {\em location-based} and user to user {\em anonymous} microblogging services complement classical social media platforms and their design differences open the question if classical observations on information spreading are still applicable.
One emerging platform in this regard is the {\em Jodel} mobile-only microblogging app.
Launched in 2014, it has been widely adopted in several European countries and Saudi Arabia.
Like Twitter, Jodel enables users to share short posts of up to 250 characters long and images, \ie microblogging.
Unlike Twitter and other classical social media platforms, Jodel {\em i)}~does not have user profiles rendering user to user communication anonymous, and {\em ii)} displays content only in the proximity of the user's location, thereby forming local communities.
Despite the emerging use of such platforms, little is known on how their key design differences impact information propagation. %

In this paper, we present the first study on information spreading in such an emerging platform by investigating the hashtag propagation in Jodel as a prominent application in this space.
We take a detailed look on hashtag propagation through the lens of a platform operator by having the unique opportunity to analyze data provided by Jodel for messages posted in Germany from September 2014 to August 2017.
This longitudinal data set enables us to study how this key design pattern of forming local communities by only displaying content to nearby users influences the hashtag usage and compares to the global counterpart Twitter.
Our study is based on using established metrics designed to capture the spatial focus and spread of Twitter hashtags~\cite{spatiotemp} to Jodel.
We show that these metrics can be applied to the temporal dimension to cover the spread of hashtags in time, enabled by our longitudinal observation period.
We further study similarities in hashtag usage between cities and their spacial impact---finding that larger cities/communities influence the smaller ones.
The correlation of spatial and temporal metrics reveal that hashtags can be grouped into four different hashtag classes distinguished by their spatial and temporal extent.
In the last step we show that these groups are distinguishable by machine learning models, informed by manual labeling of 450 most frequently used hashtags.
Our main contributions are as follows:
\begin{itemize}[noitemsep,topsep=5pt,leftmargin=9pt]
	\item We provide the first comprehensive study of hashtag usage in a local user to user anonymous messaging app.
	We find that Jodel's popular hashtags are used country-wide, whereas less popular hashtags tend to be more local.

	\item We show that classical metrics capturing the spatial propagation can be applied to the temporal domain.
	By applying these metrics, we see that popular hashtags are used over the long-run, while less popular hashtags tend to be more short-lived.

	\item We show that the used hashtags can be grouped into four classes by their spatial and temporal extent.
	We further show that these four groups can be learned by statistical models with high accuracy, based on comparing five different classifiers (k-nearest neighbour, regression trees, naive bayes, LDA, ZeroR).
	Thus, statistical methods can distinguish between different meme types found in Jodel.
\end{itemize}

\clearpage
\afblock{Paper structure.}
We introduce Jodel in Section~\ref{sec:jodelapp} and discuss related work in Section~\ref{sec:related_work}.
Section~\ref{sec:Dataset_Description_and_Statistics} introduces our Jodel dataset to which we apply established hashtag propagation metrics in Section~\ref{sec:hashtag_usage_in_jodel}.
In Section~\ref{sec:hashtag_classification}, we show that our findings can be leveraged to classify hashtags automatically.
We conclude the paper in Section~\ref{sec:future_work_and_conclusion}.

\begin{figure}[t]
	\centering
	\includegraphics[width=0.99\linewidth]{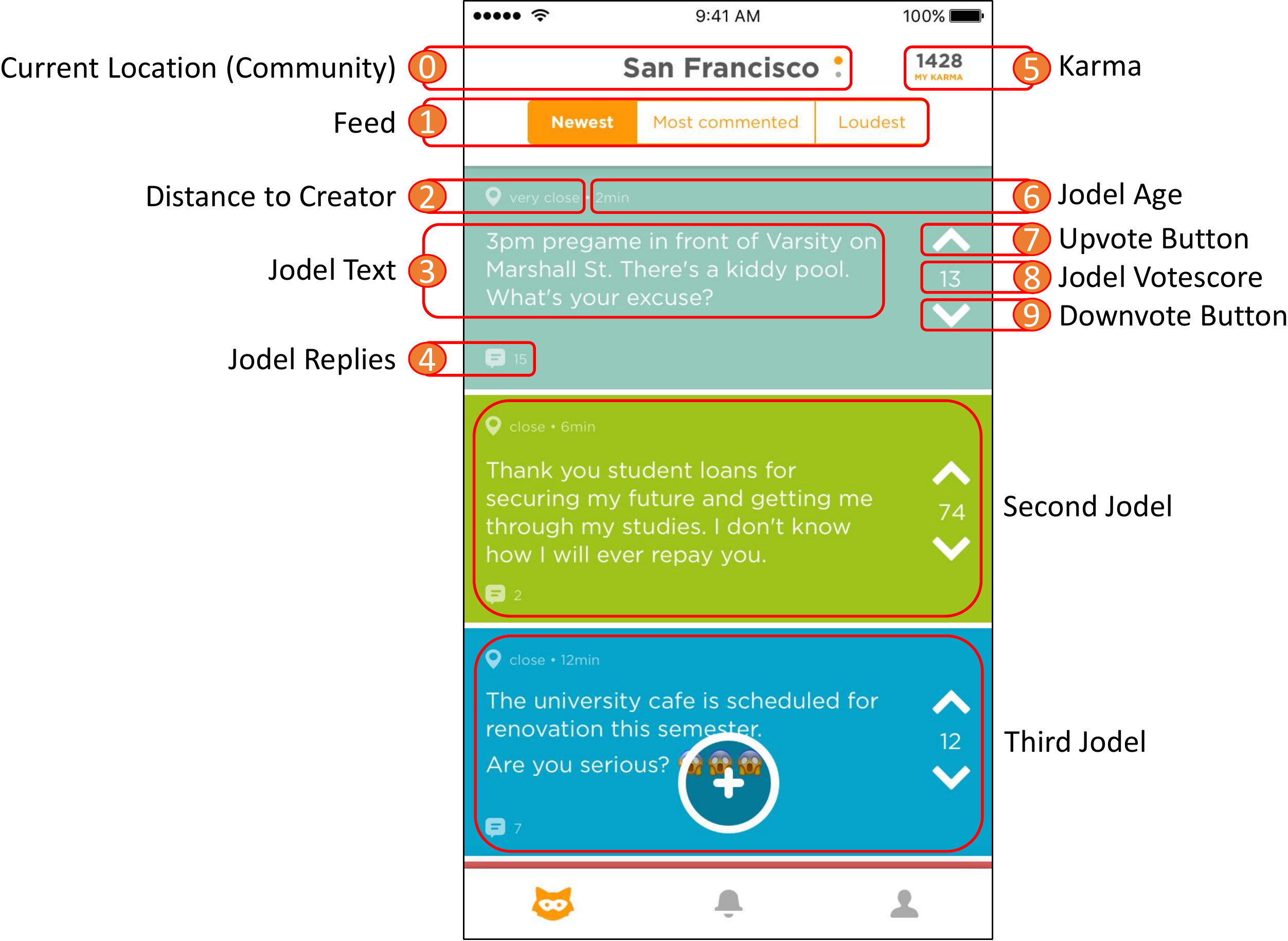}
	\caption{
		\textbf{Jodel iOS mobile application.}
	}
	\label{fig:jodelapp}
    \vspace*{-.5cm}
\end{figure}

\section{Jodel - Local Messaging App}
	\label{sec:jodelapp}

	Jodel\footnote{Jodel, German for yodeling, a form of singing or calling.}
	is a mobile-only messaging application (main-screen shown in Figure~\ref{fig:jodelapp}).
	Unlike classical social media apps, it is location-based and establishes local communities to the users' location \protect\encircle{0}.
	Within these communities, users can post both images and textual content of up to \numberNumCharactersPerJodel{} characters length \protect\encircle{3}\, (\ie micro\-blogging) {\em anonymous} to other users---and reply to posts forming discussion threads \protect\encircle{4}.
	Posted content is referred to as ``Jodels'' and are colored randomly \protect\encircle{3}.
	These posts are only displayed to other users within close (up to $20\,\text{km}$) geographic proximity \protect\encircle{2}.
	This ability to only consume local content is absent in typical social networks (\eg Twitter) that enable global communication and thus makes the study of information spread interesting.

	All communication is {\em anonymous to other users} since no user handles or other user-related information are displayed.
	Only {\em within} a single discussion thread, users are enumerated and represented by an ascending number in their post order. %
	There are three different content feeds \protect\encircle{1}: \emph{i) newest} showing the most recent threads, \emph{ii) most commented} showing the most discussed threads and \emph{iii) loudest} showing threads with the highest voting score (cf. later).
	Additionally, users can subscribe to thematic channels.
	Each post can contain hashtags and the app enables to display further local posts with the same hashtag by clicking on a hashtag in a post.

	Jodel employs a community-driven filtering and moderation scheme to avoid adverse content.
	For any social network or messaging app, community moderation is a key success parameter to prevent harmful or abusive content.
	The downfall of the Jodel-alike YikYak anonymous messaging application highlighted that unsuccessfully preventing adverse content can seriously harm it~\cite{yikyakNyTimes}.
	In Jodel, content filtering relies on a distributed voting scheme in which every user can increase or decrease a post's vote score by up- (+1) \protect\encircle{7} or downvoting (-1) \protect\encircle{9}, \ie similar to StackOverflow.
	Posts reaching a cumulative vote score \protect\encircle{8} below a negative threshold \mbox{(\eg -5)} are not displayed anymore.
	Depending on the number of vote-contributions, this scheme filters out bad content while also potentially preferring mainstream content.
	As a second line of defense, Jodel employs community moderators who decide on removing reported posts.

\section{Related Work}
\label{sec:related_work}

Our paper relates to three main areas within research: \emph{i)} general meme spread modelling, \emph{ii)} the use case of microblogging, \eg{} Twitter, and \emph{ii)} others; which we will discuss next.

\afblock{Spreading \& contagion models.}
A classical approach to study information diffusion is applying spreading models.
Epidemic models have been applied to memes, where a meme can \emph{infect} people by coming in contact with it (SIR models)---possibly extended with mechanics for \emph{recovery} (SIRS models), \eg in \cite{epidemics_2,epidemics_3}. %
Although these approaches model the growth of hashtag popularity well, most fail to map the typical power-law decay \cite{model_1}.
Their application to hashtags is further limited by requiring an infection time, \ie{} when a user learns about a hashtag.
Passive information consumption such as reading is typically not included in most social network data.

\afblock{Twitter.}
The study of hashtag usage and diffusion mostly targets Twitter given its popular use of hashtags and ability to geotag posts.
Although Twitter has no boundaries regarding distance (\ie unlike Jodel), cities closer to each other share more hashtags, supported by an analysis of the Twitter trending topics in \cite{butterflies}.
The authors find three clusters of hashtag similarity across the biggest cities in the US and speculate that the spread is related to airports.
To study non-stationary time series of hashtag popularity, \cite{local_variation} applies a statistical measure originally used for neuron spike trains to hashtags.
It is capable of giving information on how regularly hashtags are used.
They find that low to mediocre popular Twitter hashtags are on average rather bursty, while extremely popular ones are posted more regularly.
The influence of content (\eg politics, music, or sports) on the hashtag adoption is studied in \cite{topics}.
The authors find that especially political hashtags are more likely to be adopted by a user after repeated exposure to it than hashtags of other topics.

To capture the spatio-temporal dynamics of Twitter hashtags, {\em focus}, {\em entropy}, and {\em spread} were proposed as metrics~\cite{spatiotemp}.
By applying these metrics to Twitter, the authors find hashtags to be a global phenomenon but the distance between locations to constraint their adoption.
We will use these metrics to study Jodel and we extend them with a temporal dimension within our analysis.
To study the how cities impact each other regarding hashtag adoption, \cite{spatiotemp} also proposed a spatial impact metric to capture the similarity of hashtag uses in two cities---a metric that we will adopt likewise.
They show that the biggest influencers were big cities with large user bases.

\afblock{Other platforms.}
Besides Twitter, few studies consider other platforms.
The sharing cascades in Facebook are studied in \cite{facebook}.
Similar cascades are found by studying how the blogosphere and the news media influence each other~\cite{blogosphere}.
Memes do not have to be in the form of images or text, but can also be videos--as such, \eg \cite{youtube} studies the diffusion of memes on Youtube.

Other works focused on the influence of events in terms of the spreading behavior.
\Eg \cite{class_2,class_3} used statistical classifiers on contextual features to distinguish between memes and events. %
Researchers have also tried to detect events, \eg by analyzing the Twitter stream \cite{detection_1, detection_2} and inferring where an event happens \cite{detection_3}.
There were also efforts to detect earthquakes and estimating the epicenter in realtime \cite{earthquake}.
Also, user positions can be at least vaguely estimated as shown in \cite{user_estimation}.

We complement these works by studying the hashtag usage and diffusion on Jodel.
Its property to only display posted content to nearby users differentiates Jodel from other studied social networks that disseminate content globally (\eg{} Twitter or Facebook).
It thus {\em might}---and as we will see: \emph{will}---feature a fundamentally different spreading behavior.

\begin{table}[t]
	\small
	\centering
	\begin{tabular}{|l|r|l|} %
		\hline
		\textbf{Metric}		& \textbf{\#Entries}	& \textbf{Description}\\ \hline\hline
		Hashtag Uses		& $41,038,733$			& \# of hashtags occurrences\\\hline
		Hashtags         	& $13,110,573$			& \# of different hashtags\\\hline
		         	        & $11,092,360$			& \# of hashtags used only once\\\hline
		Messages			& $26,955,008$			& \# of messages that contained hashtags\\\hline
		Users				& $1,240,404$			& \# of users posting contents with hashtags\\\hline
		Locations			& $6,830$				& \# of different posting locations/cities\\\hline
	\end{tabular}
	\caption{
		\textbf{Hashtag dataset statistics.}
		The data ranges from the application start in late 09/2014 up to beginning of 08/2017.
	}
	\label{tab:dataset}
    \vspace*{-1cm}
\end{table}

\section{Dataset Description and Statistics}
\label{sec:Dataset_Description_and_Statistics}
	The Jodel network operator provided us with {\em anonymized} data of their network.
	This obtained data contains post, user and interaction metadata and message contents created within Germany only.
    It spans multiple years from the beginning in September 2014 of the network up to August 2017.
	The dataset only includes infromation users have publicly posted and thus visible to all other Jodel users.
	Structurally, our available dataset is built up from three object categories: interactions (about 400\,M records), content (about 285\,M records), and users (about 900\,k records).
	The location of each post (and thus each hashtag) is available on a {\em city-level granularity}. %

	\afblock{Hashtags.}
	We have extracted hashtags from the message contents by applying a regular expression matching a `\#' followed by any amount of alphanumeric characters (including German umlauts and Eszett), dots, dashes or underscores.
	This resulted in a total amount of about $41\,\text{M}$ hashtag uses within $26\,\text{M}$ different messages and $13\,\text{M}$ different hashtags.
    These messages where created by $1.2\,\text{M}$ users having posted in about $7\,\text{k}$ different locations.

    Within the set of hashtags, we observe that $11.1\,\text{M}$ are only used once.
    This leaves about $2\,\text{M}$ hashtags that have been used multiple times, \ie{} $\geq2$, and therefore are suited for our hashtag propagation analysis at all.
    After manual sample screening, the predominant reason for this huge amount of hashtags occurring only once is that on Jodel, they are often used as a unique stylistic feature, support content, or are misspelled reuses---in contrast to a self-categorization that might be expected.

\section{Jodel Hashtag Usage and Spread}
\label{sec:hashtag_usage_in_jodel}
In this section, we analyze the spread and propagation of content in Jodel by using hashtags as a proxy measure.
That is, we leverage the user's ability to tag posts with hashtags to relate to topics, add categories or metadata to posts.
Although hashtags are sometimes used as a rather stylistic feature (\eg by using numbers as hashtags to link multiple character limited posts together), more popular ones overall reasonably capture topics and memes in the posts.
We will see that some hashtags are specific to the Jodel platform and very local possibly due to its location-based design.
Beginning our analysis in this Section with a study of hashtag popularity, we follow this up with their spatial and temporal spreading extent.
We lastly study the hashtag usage in different cities and how they influence the hashtag adoption.

\subsection{Overall Hashtag Use}
Our data set consists of $27\,\text{M}$ posts with hashtags.
We overall find $41\,\text{M}$ occurrences of $13\,\text{M}$ unique hashtags of which only $2\,\text{M}$ are used multiple times (cf. Table~\ref{tab:dataset}).

\begin{figure}[t]
	\centering
	\begin{subfigure}[c]{0.48\columnwidth}
		\includegraphics[width=1\linewidth]{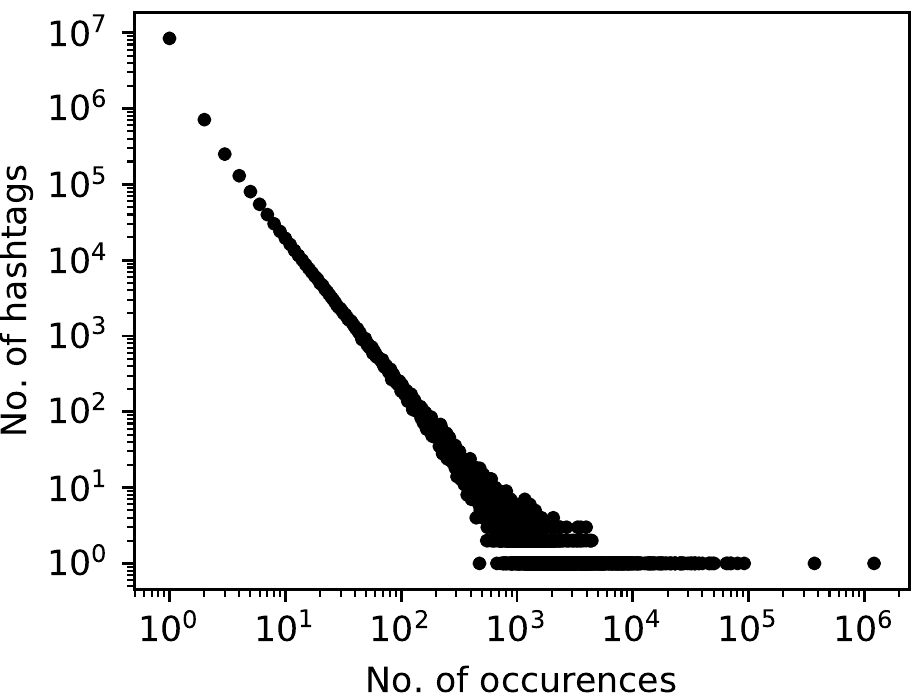}
		\subcaption{Hashtag distribution}
		\label{hashtags_meta_occurences}
	\end{subfigure}
	\begin{subfigure}[c]{0.48\columnwidth}
		\includegraphics[width=1\linewidth]{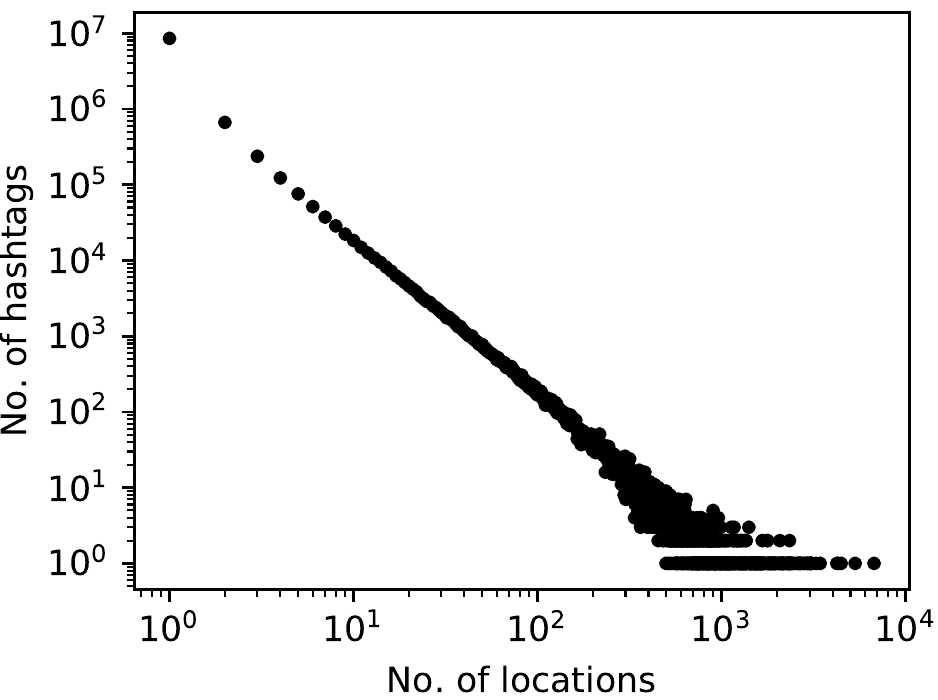}
		\subcaption{Location distribution}
		\label{hashtags_meta_locations}
	\end{subfigure}
	\caption{These figures show \emph{a)} the hashtag distribution w.r.t occurrences and the corresponding amount, \emph{b)} the location distribution w.r.t occurrences for a hashtag and the corresponding amount---both distributions are heavy-tailed.
	}
	\label{hashtags_meta}
    \vspace*{-.5cm}
\end{figure}

\afblock{Popularity.}
We begin by studying the hashtag popularity.
Figure~\ref{hashtags_meta_occurences} shows the distribution of a hashtag's occurrence (x-axis) vs. the corresponding amount of unique hashtags in our dataset (y-axis) on a log-log scale.
We observe that the vast majority of hashtags are only used few times. %
The distribution is heavy-tailed and of similar shape, as observed in Twitter~\cite{spatiotemp}.

\afblock{Location distribution.}
We next study how many hashtags (y-axis) are used in how many locations (x-axis) in Figure~\ref{hashtags_meta_locations}.
We see that not only the occurrences per hashtags is heavy-tailed, but also their geographic spread.
These results are also very similar to Twitter~\cite{spatiotemp}.

\afblock{Findings.}
We find most hashtags are being used only very few times.
The hashtag usage follows a heavy-tailed distribution, which also holds true for the number of different locations in which they occur.
That is, only a few hashtags are heavily popular and used in many locations---others to a lesser extent, or not.

\begin{figure*}[t]
	\centering
	\begin{subfigure}[c]{0.39\linewidth}
        \hspace*{-.5cm}
		\centering
		\includegraphics[height=3.6cm]{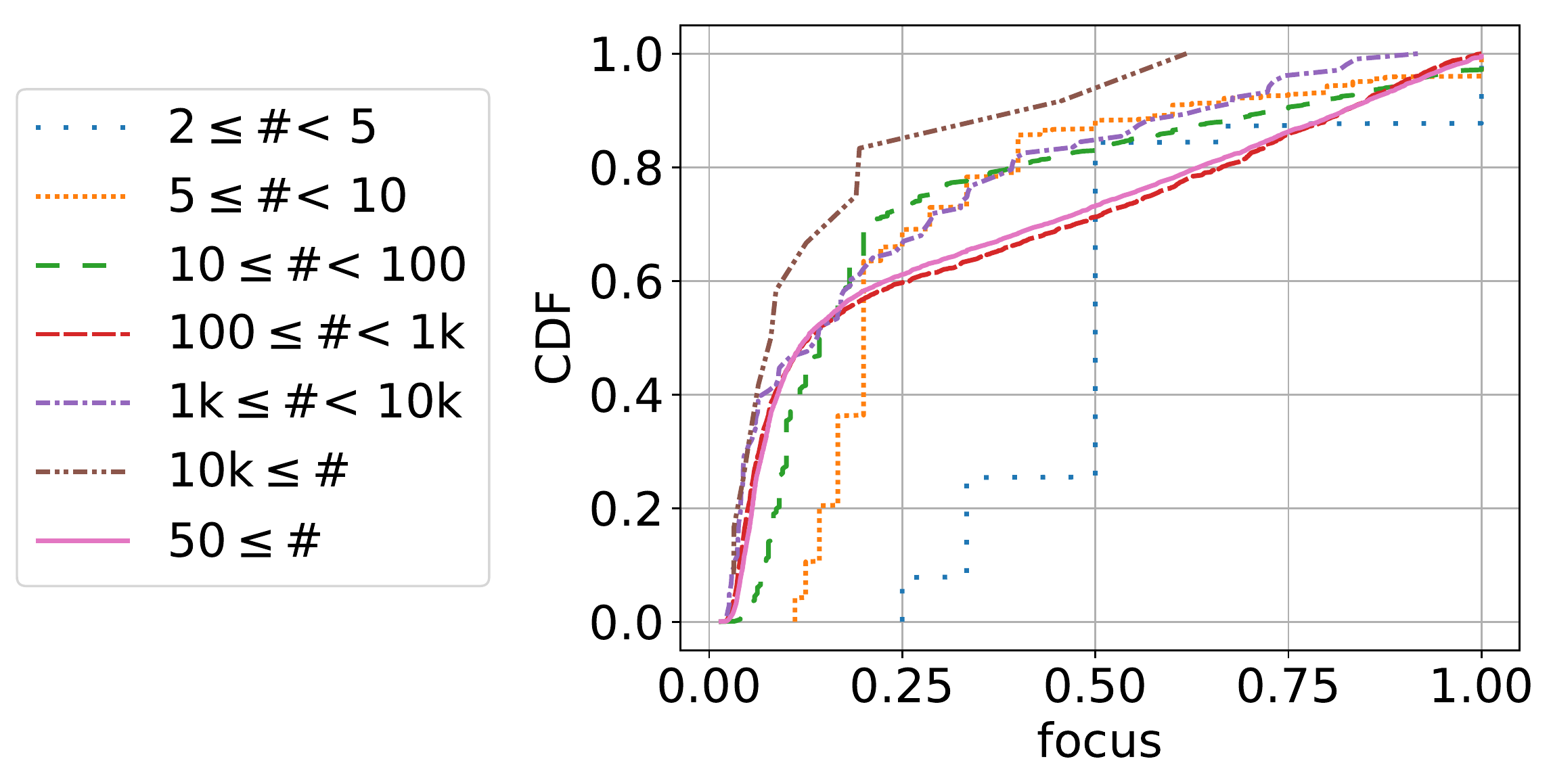}
		\subcaption{Spatial focus CDF}
		\label{fig:focus_cdf}
	\end{subfigure}
	\begin{subfigure}[c]{0.27\linewidth}
		\includegraphics[height=3.6cm]{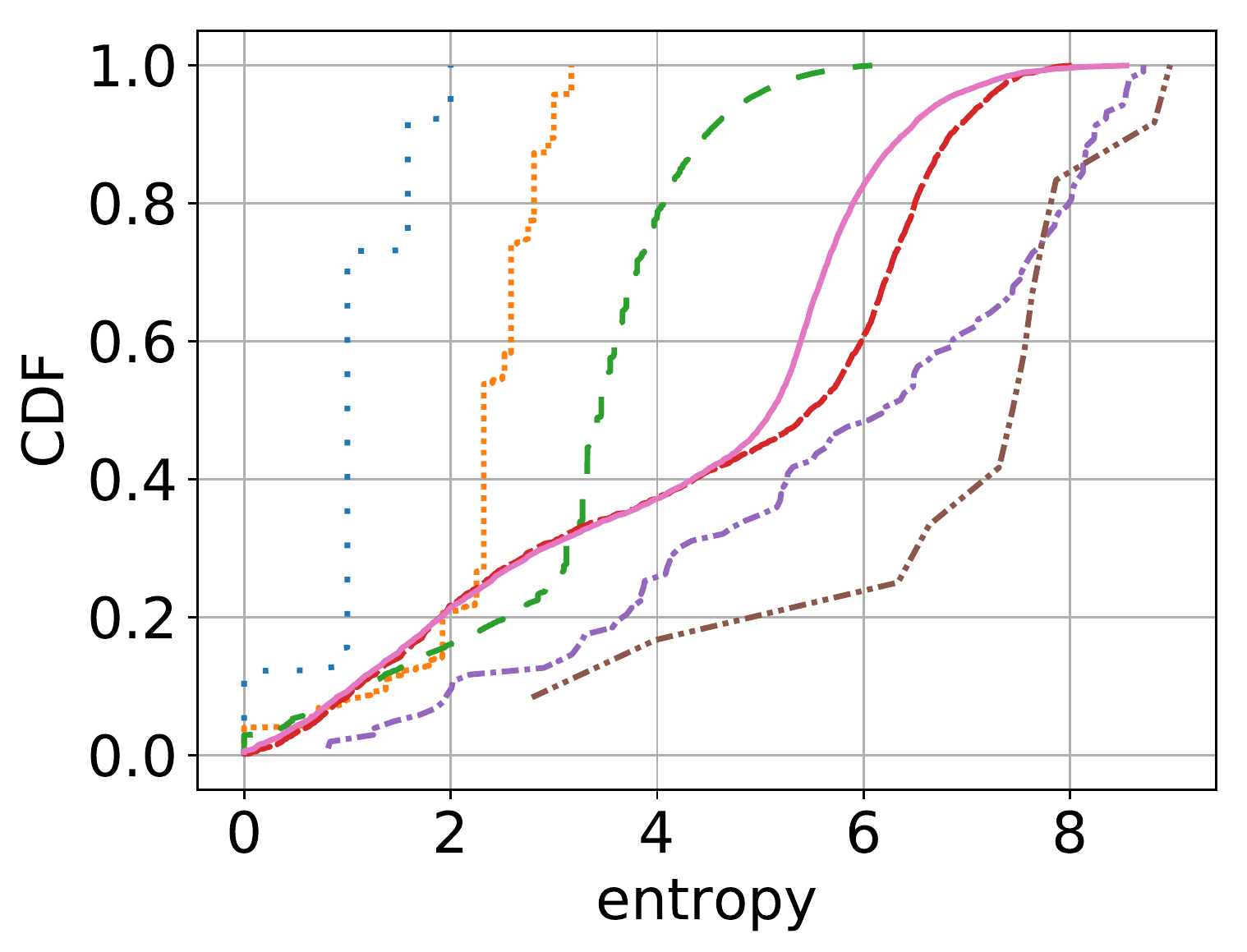}
		\subcaption{Entropy CDF}
	\label{fig:entropy_cdf}
	\end{subfigure}
	\begin{subfigure}[c]{0.24\linewidth}
		\includegraphics[height=3.6cm]{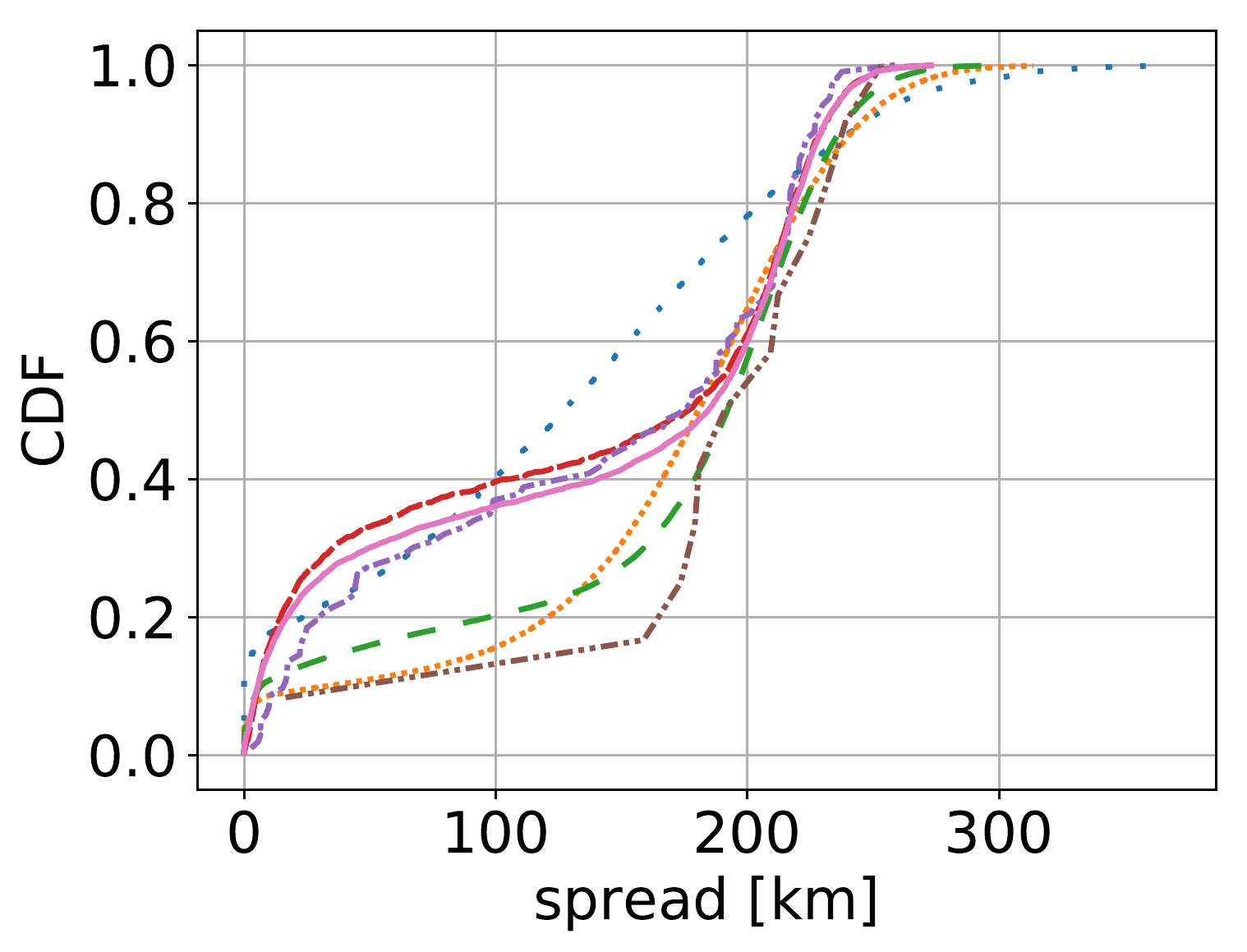}
		\subcaption{Spread CDF}
		\label{fig:spreadf_cdf}
	\end{subfigure}
    \vspace*{-.25cm}
	\caption{Spatial hashtag metrics: focus, entropy and spread (left to right).
        All figures follow the partitioning by hashtag occurrences shown left.
        \emph{a)} The more popular a hashtag is, the more unfocused it gets. Most hashtags show a very low focus, while very unpopular hashtags naturally tend to be more focused.
        \emph{b)} Likewise, less used hashtags naturally can only be used in few locations, while more popular hashtags are used in many locations.
        \emph{c)} Some hashtags on average span geographically only up to $\pm50\,\text{km}$, whereas most are used all over the country.
    }
    \label{fig:spatial_metrics}
    \vspace*{-.5cm}
\end{figure*}

\subsection{Spatial Properties of Jodel Hashtags}
\label{sec:spatial_metics}
We next study spatial properties of Jodel hashtags, \eg if a certain hashtag only occurs in a local community or over which geographic distance the usage of a countrywide hashtag is spread.
To capture these spatial properties, we use three hashtag metrics originally proposed for Twitter: \emph{focus}, \emph{entropy}, and \emph{spread}~\cite{spatiotemp}.
These metrics enable us to judge if content diffusion in Jodel actually is---due to its design---indeed {\em more local} than a comparable microblogging platform without geographical communities, like, \eg Twitter.

\afblock{Data filtering.}
We restrict our set of hashtags by only considering hashtags that occurred first in 2016 or later.
This way, we focus on a time in which the app has an established user base in Germany.

\afblock{Focus.}
The focus metric captures how locally or globally (\ie in our case countrywide) focused the use of a hashtag is~\cite{spatiotemp}.
To achieve this, the set of hashtags and the set of locations are defined as $H$ and $L$, respectively, of which for a given hashtag $h \in H$ and location $l \in L$, $O^h_l$ is the set of occurrences of $h$ in $l$.
Then, the probability of observing a hashtag $h$ in a location $l$ is defined as:
$$P^h_l = \frac{\vert O^h_l \vert}{\sum_{m \in L}\vert O^h_m \vert}$$
The focus location of a hashtag is defined as the location with most occurrences of that hashtag and further provides a fraction of the occurrences in the focus location compared to the number of overall occurrences.
It is defined as $ F^h = \max\limits_{l \in L} P^h_l$.
Then, the focus for hashtag $h$ is defined as a tuple of the focus location $l_f=F^h$ and its probability $P^h_{l_f}$.
Hashtags only popular in a few cities will have a higher focus, whereas globally popular hashtags will have a lower one.
A limitation of the focus metric is that it provides information only about one single location, but nothing about the distribution.

We show the focus distribution of hashtags in Figure~\ref{fig:focus_cdf}, where a series represents a CDF for a set of hashtags partitioned by their occurrence. %
As the hashtags are subject to popularity, \ie usage frequency, these partitions define different log-based groups within out dataset (cf. Figure~\ref{hashtags_meta_occurences}).
Our observation is that the focus distribution is skewed towards low focus values regardless of hashtag occurrences.
That is, 60\% of all hashtags that occur $\ge 5$ times have a focus of $\le 0.25$.
This means that from all occurrences of such a hashtag, only $25\,\%$ occur in its most popular city, whereas the remaining $75\,\%$ of the hashtag occurrences is in other cities.
Therefore, the focus distribution indicates that the usage of most hashtags is not focused on a single city but is rather spread over multiple cites.
Further, the observed skew within the distributions towards low focus values differs from hashtag usage observations in Twitter in which the hashtags' focus was uniformly distributed~\cite{spatiotemp}. %
The prevalence of low focus values is unexpected and interesting; the design of the App to only display nearby posts could have caused a skew towards high focus values, in which the usage of most hashtags would be more concentrated.
This, however, is not the case.

\afblock{Entropy.}
The entropy metric captures in how many locations a hashtag is used~\cite{spatiotemp}.
For a hashtag $h$, it is defined as:
$$ E^h = -\sum_{l \in L} P^h_l \log_2 P^h_l $$
This metric defines the minimum number of bits required to represent the amount of a hashtag's locations it has spread to.
The higher the diffusion of a hashtag, the higher its entropy; \ie the entropy defines the number of locations a hashtag occurred in by the power of $2$.
For more often used hashtags, both entropy and focus are resistant to small changes in the data (\eg single occurrences in another ten locations).

Similar to the focus, we show the entropy distribution as CDFs for hashtags likewise partitioned by occurrences in Figure~\ref{fig:entropy_cdf}.
We observe that only a negligible number of hashtags is used in a single city (entropy 0).
Looking into the different partitions, we identify that less popular hashtags clearly tend to a smaller entropy.
However, for the more popular hashtags having at least 50 occurrences, more than $60\,\%$ of the hashtag occurrences are in $\ge 16$ cities (entropy 4).
As already indicated by the focus distribution, the usage of most hashtags is thus not concentrated to a single city only but spread over multiple cities.
In summary, the hashtag usage shows a trend to higher entropy values with an increased number of occurrences; the more popular a hashtag is, there more it is spread across different cities, which supports our findings for the focus.

\begin{figure*}[ht]
	\centering
	\begin{subfigure}[c]{0.39\linewidth}
        \hspace*{-.5cm}
		\includegraphics[height=3.6cm]{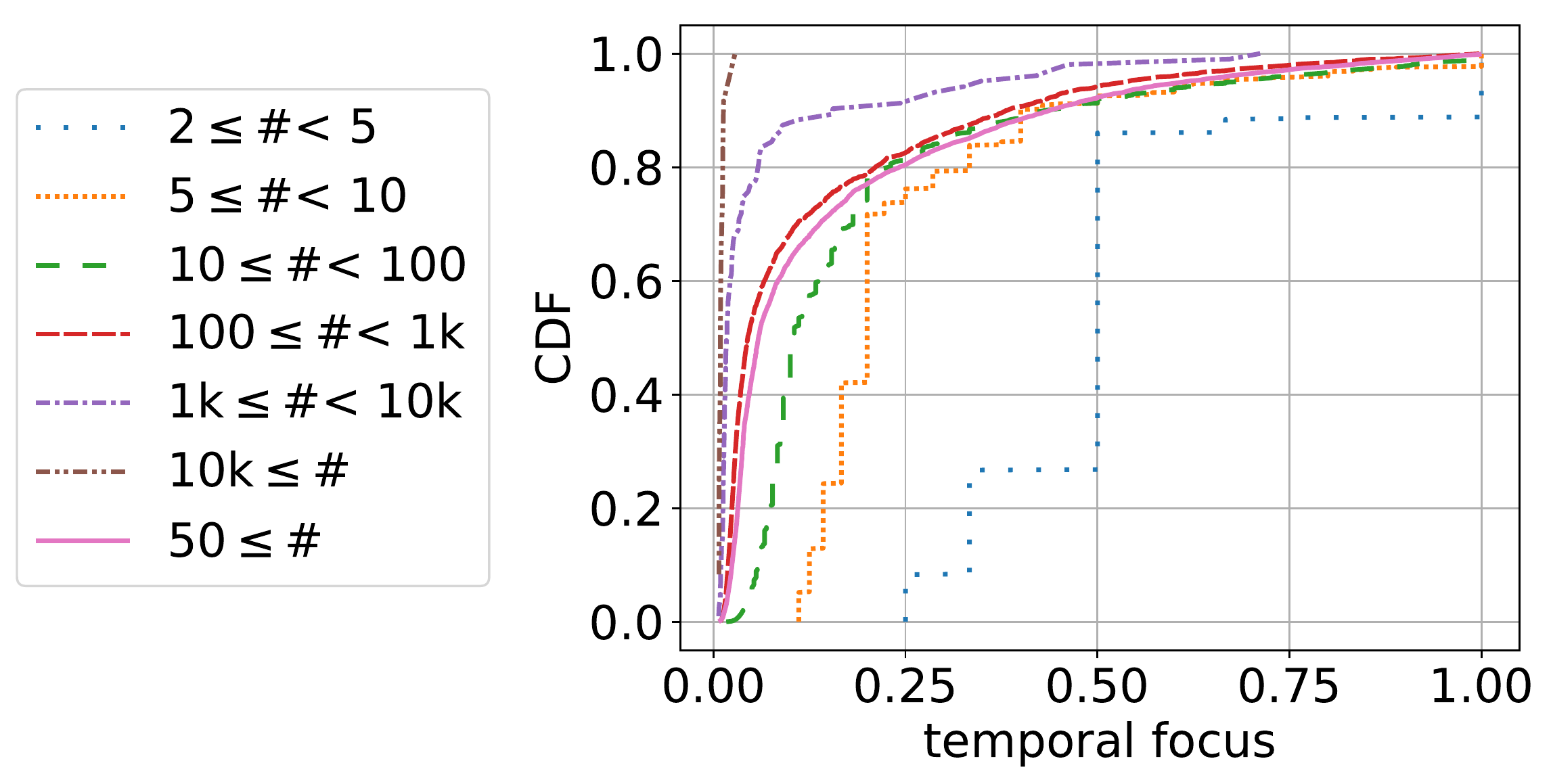}
		\subcaption{Temporal Focus CDF}
		\label{fig:temporal_focus_cdf}
	\end{subfigure}
	\begin{subfigure}[c]{0.27\linewidth}
		\includegraphics[height=3.6cm]{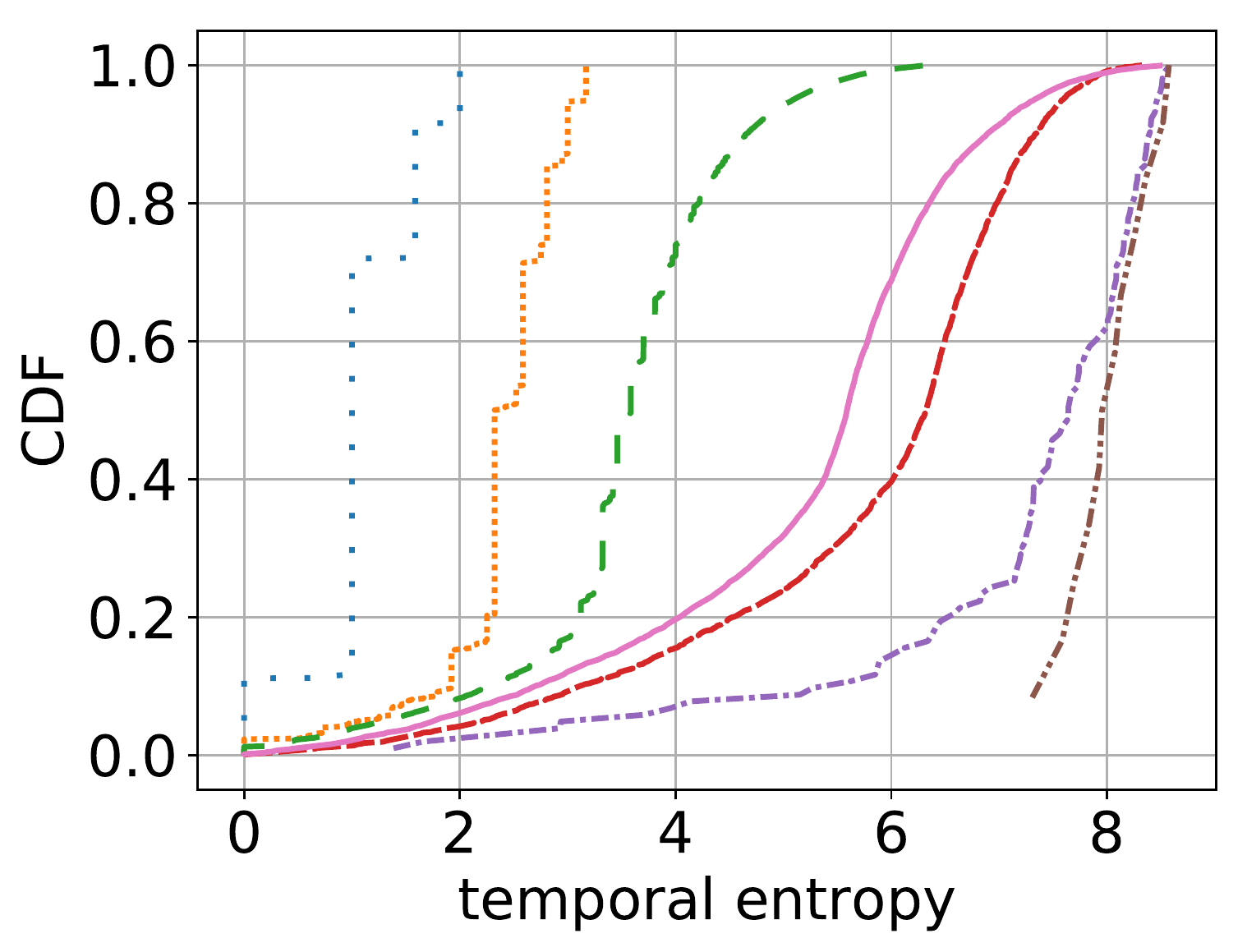}
		\subcaption{Temporal Entropy CDF}
		\label{fig:temporal_entropy_cdf}
	\end{subfigure}
	\begin{subfigure}[c]{0.24\linewidth}
		\includegraphics[height=3.6cm]{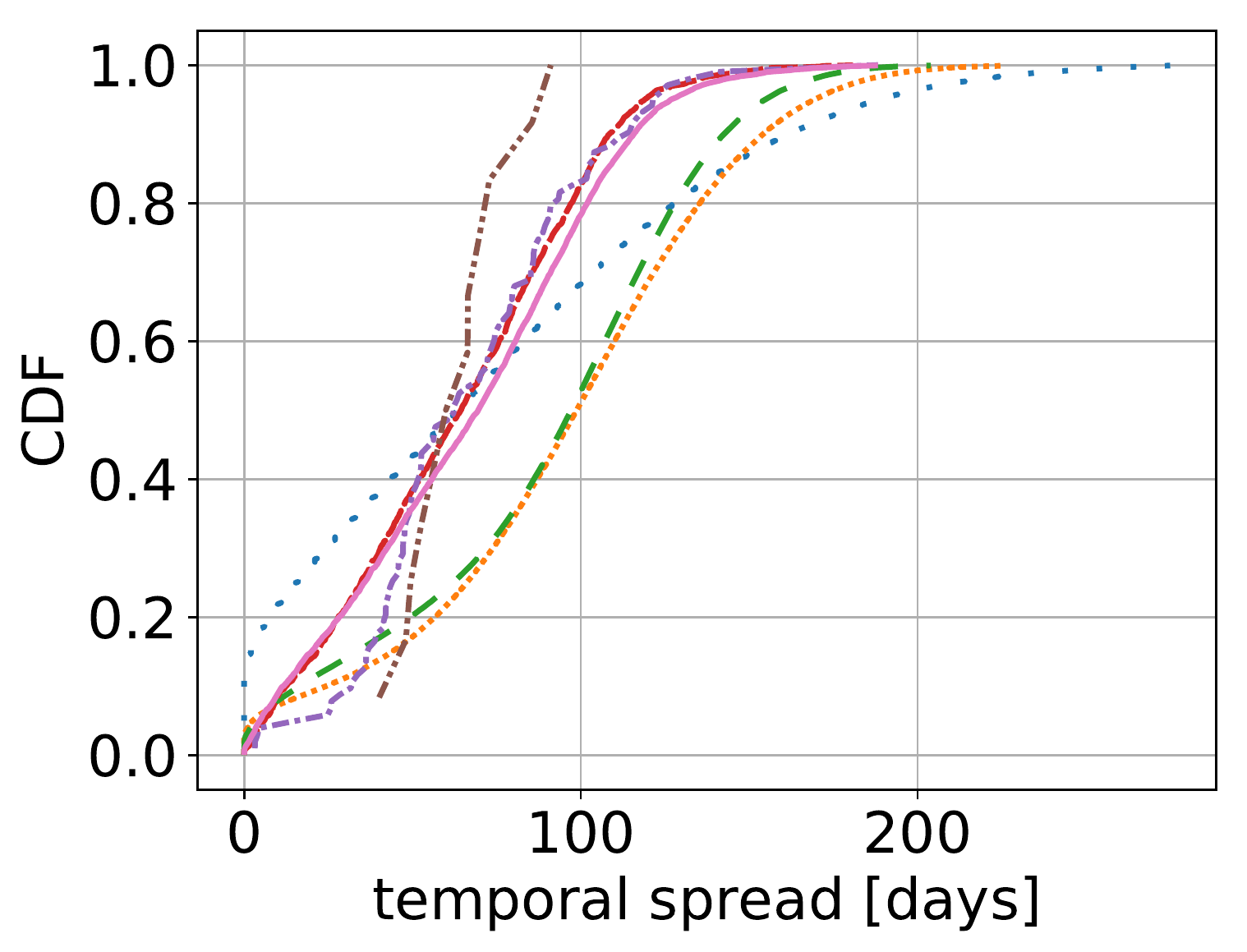}
		\subcaption{Temporal Spread CDF}
		\label{fig:temporal_spreadf_cdf}
	\end{subfigure}
    \vspace*{-.25cm}
	\caption{
		Temporal hashtag metrics: temporal focus, temporal entropy, and temporal spread (left to right) are temporal adoption to the spatial counterparts (cf. Section~\ref{sec:spatial_metics}).
        All figures follow the partitioning by hashtag occurrences shown left.
		\emph{a)} The temporal focus decreases with hashtag popularity, \ie they become used over longer time periods.
		\emph{b)} This finding is supported by the temporal entropy showing that more popular hashtags are more widespread across multiple dates.
		\emph{c)} The temporal spread indicates a possible distinction between a smaller set of short-lived hashtags and a large set of long-lived hashtags.
	}
	\label{temporal_metrics}
    \vspace*{-.5cm}
\end{figure*}

\afblock{Spread.}
To obtain information about the geographical expansion, we can use the spread metric defined as the mean distance of the geographic midpoint of the set of hashtag occurrences~\cite{spatiotemp}:
\vspace*{-5px}
$$ S^h = \frac{1}{\vert O^h \vert} \sum_{o \in O^h} D(o, G(O^h)) $$
where $D$ is the distance in kilometers and $G$ is the weighted geographic midpoint.
As on our scale (Germany), the spherical shape of the Earth is only of minor importance, we use the weighted average latitude and longitude as the midpoint.
A spread of 50\,km thus means that the average usage of a hashtag occurs within $\pm 50$\,km.

We show the spread distribution again as CDFs of partitions by occurrences in Figure~\ref{fig:spreadf_cdf}.
The distributions reveal that there are three groups of hashtags:
\emph{i)} Only rarely used hashtags ($\le 5$ occurrences) show a rather linear spread,
\emph{ii)} More frequently used hashtags ($5..100$ occurrences) show a slight bimodal distribution as they either have a small spread up to $50\,\text{km}$, or most of them show a rather big spread $>150\text{km}$.
The same holds true for hashtags that are heavily used.
\emph{iii)} Hashtags that are used often, but do not belong to the heavy tail, strengthen the bimodal observation as about $40\,\%$ only have an up to $50\,\text{km}$, whereas most others are spread wider.

We note that higher spreads are likely the value a Germany-wide hashtag may achieve.
While there is no (known) comparable analysis for Twitter or similar platforms, we conclude that the lower-spread hashtags are most probably an implication of Jodel's nature building location-based communities.
\Ie there are hashtags that are used in a geographically restricted area at small distances.

\afblock{Findings.}
We observe that most hashtags in Jodel are used rather countrywide, \ie their usage does not concentrate on single cities and spreads over larger geographic distances.
This is unexpected since the design of Jodel to form local geographic communities could also result in a more geographically focused usage of hashtags.
However, while most hashtags are used rather globally, up to $40\,\%$ have a local spread of $\pm 50$\,km and thus are a potential consequence of Jodels' design.

\afblock{Twitter Comparison.}
A direct comparison to \cite{spatiotemp} can be made within our series of hashtags at least having 50 occurrences (pink solid lines).
While the \emph{focus} CDF for Twitter hashtags is rather linear with the exception of $20\,\%$ having focus $1$, the focus on Jodel is distributed in an opposite fashion.
That is, $60\,\%$ of Jodel hashtags ($\geq50$ occurrences) tend to be non-focused below a value of $0.25$, but are likewise equally distributed above---having almost no hashtags with focus $1$.
As for the \emph{entropy}, most hashtags on Twitter are used very locally, which can only be observed for least popular hashtags on Jodel---many more popular hashtags are used across the country.
Similarly, the \emph{spread} on Twitter is either local for few hashtags, but then increases linearly, which is identical for the least and heavily popular hashtags on Jodel---others show a pronounced bimodal distribution between local and countrywide scope.

\subsection{Temporal Properties of Jodel Hashtags}
We are next interested in studying how hashtags develop over time (\eg gain in popularity).
This is possible given our longitudinal data set.
Therefore, we adopted focus, entropy, and spread for our temporal analysis.
Instead of locations as in our spatial analysis, we use the creation time of a hashtag's post (grouped to days for focus and entropy) for each hashtag occurrence.
The grouping to days makes sense due to limited content presence within the usually highly dynamic Jodel feeds for larger communities.

\afblock{Temporal Focus.}
We show the temporal focus distribution as CDFs partitioned by hashtag occurrences in Figure~\ref{fig:temporal_focus_cdf}.
Recall that the temporal focus now defines the probability of a hashtag to be used on its most popular day, \ie a temporal focus of 1 indicates that a hashtag is exclusively used on a single day whereas a focus of near 0 would suggest a spread over the entire observation period.
We observe that about $80\,\%$ hashtags have a low temporal focus $\leq0.25$, suggesting that their lifetime is not focused on a single point in time.
The more popular they become, the temporal focus decreases, \ie they remain popular over time.
However, least popular hashtags tend to a higher temporal focus in comparison.
In summary, there are almost no hashtags focused to a single day.
For those that are being used only a few times, this implicates random re-use that is probably not correlated, whereas popular hashtags are used throughout the observation period.

\afblock{Temporal Entropy.}
The temporal entropy defines the number of days on which a hashtag is used.
We show its distribution as CDFs partitioned by hashtags occurrences in Figure~\ref{fig:temporal_entropy_cdf}.
We observe that only a negligible amount of hashtags are used on exactly one day (entropy 0).
Except for the only rarely used hashtags, more than $90\,\%$ occurrences have an entropy above $2$, \ie they were used on more than $4$ ($2^2$) days.
Further, the higher the occurrences (popularity) of a hashtag, the higher the entropy.
This indicates that popular hashtags are used for longer time periods.

\afblock{Temporal Spread.}
The temporal spread defines the average time period in days in which a hashtag is used.
For example, a temporal spread of 50\,days means that the average usage period of a hashtag is $\pm 50$\,days (past \& future) from the temporal weighted midpoint.
We show the distribution of the temporal spread as CDF again partitioned by hashtag's occurrences in Figure~\ref{fig:temporal_spreadf_cdf}.
We observe that the temporal spread is distributed equal (linear CDF) across all partitions.
However, the activity period is again influenced by the popularity of a hashtag; the more popular a hashtag is, the higher is the temporal spread.
The presented series that only include hashtags with very few uses depict a large set of hashtags with a temporal spread of more than $100\,\text{days}$---the significant skew towards a larger spread strengthens our belief that such hashtags occur independently from each other (cf. temporal focus).

\afblock{Findings.}
Popular hashtags in Jodel are seldomly a flash in the pan but are mostly used over extended time periods.
In particular, the more popular a hashtag is, the longer and frequent its usage period becomes, whereas less popular ones rather occur independently from each other.
This is interesting since the Jodel app provides---unlike Twitter---only limited functionality to search for hashtags as hashtags may only be clicked when seen in a post, \ie for a purposeful re-use it must be known. %

\subsection{Spatial vs. temporal dimensions}
\label{sec:spatial_vs_temporal}
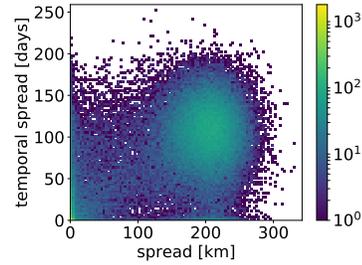
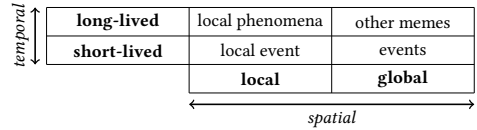
\begin{figure}[t]
	\begin{subfigure}[c]{1\linewidth}
		\centering
		\includegraphics[height=3.6cm]{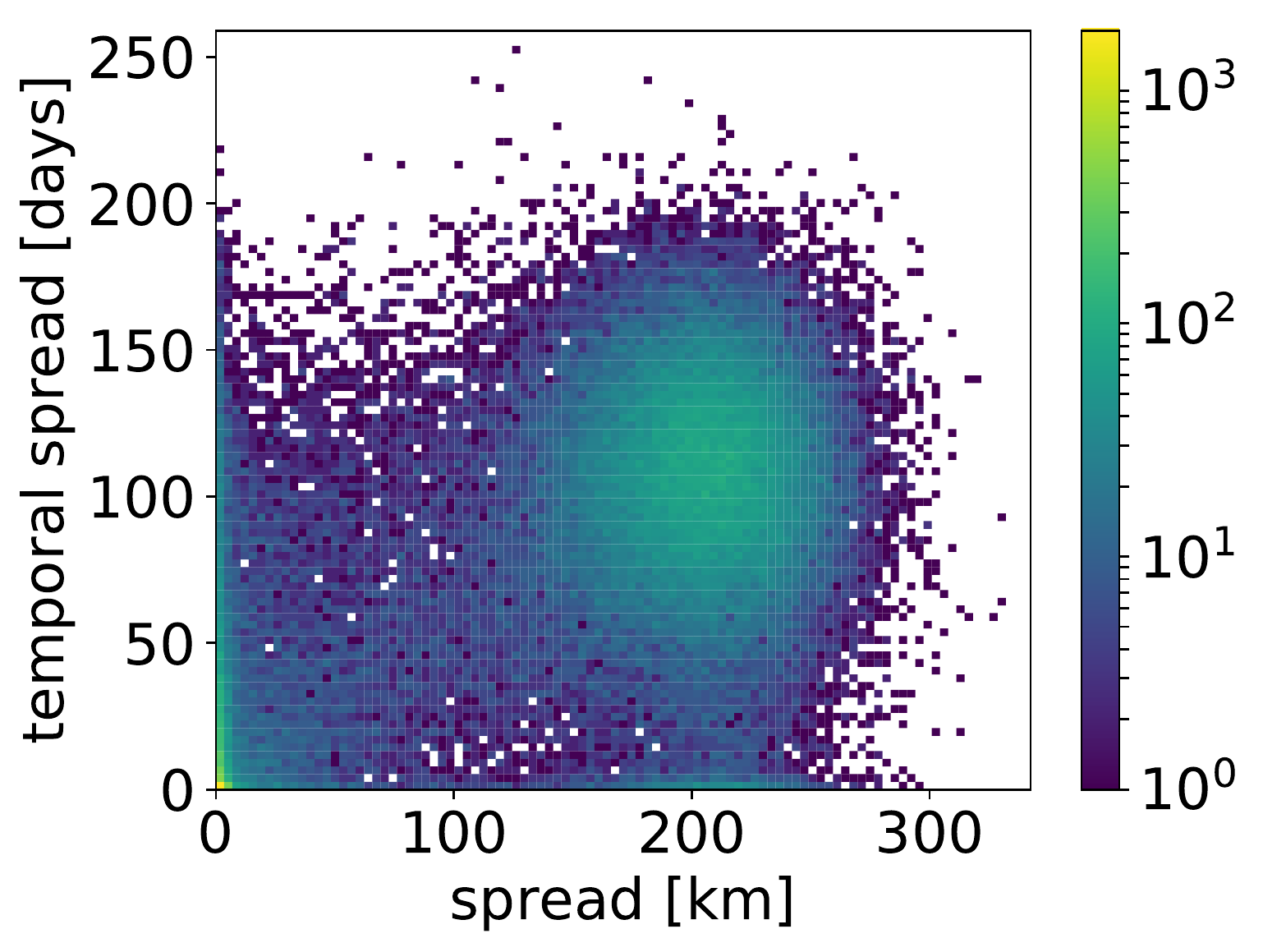}
		\caption{
            The combined spatial (x-) and temporal (y-axis) spread shown as a heatmap of hashtags occurrences (z-axis).
			We observe clusters: \emph{i)} countrywide long-lived hashtags, \emph{ii)} countrywide short-lived hashtags, and \emph{iii)} Both, local short- and long-lived hashtags.
		}
		\label{fig:spreadvstemp}
	\end{subfigure}

	\begin{subfigure}[c]{1\linewidth}
		\centering
		\resizebox{.75\linewidth}{!}{
			\begin{tikzpicture}
				\def \boxwidth {2.5}
				\def \boxheight {.5}
				\def \boxoffset {.2}

				\node (local_long) 		at (1*\boxwidth,2*\boxheight)	{local phenomena};
				\node (local_short)		at (1*\boxwidth,1*\boxheight)	{local event}; %
				\node (global_long)		at (2*\boxwidth,2*\boxheight)	{other memes};
				\node (global_short) 	at (2*\boxwidth,1*\boxheight)	{events};

				\node (desc_long) 		at (0*\boxwidth,2*\boxheight)	{\textbf{long-lived}};
				\node (desc_short)		at (0*\boxwidth,1*\boxheight)	{\textbf{short-lived}};
				\node (desc_local)		at (1*\boxwidth,0*\boxheight)	{\textbf{local}};
				\node (desc_global) 	at (2*\boxwidth,0*\boxheight)	{\textbf{global}};

				\draw (-.5*\boxwidth,2.5*\boxheight) 	-- (2.5*\boxwidth,2.5*\boxheight);
				\draw (-.5*\boxwidth,1.5*\boxheight) 	-- (2.5*\boxwidth,1.5*\boxheight);
				\draw (-.5*\boxwidth,.5*\boxheight) 	-- (2.5*\boxwidth,.5*\boxheight);
				\draw (.5*\boxwidth,-.5*\boxheight)		-- (2.5*\boxwidth,-.5*\boxheight);

				\draw (-.5*\boxwidth,2.5*\boxheight)	-- (-.5*\boxwidth,.5*\boxheight);
				\draw (.5*\boxwidth,2.5*\boxheight) 	-- (.5*\boxwidth,-.5*\boxheight);
				\draw (1.5*\boxwidth,2.5*\boxheight) 	-- (1.5*\boxwidth,-.5*\boxheight);
				\draw (2.5*\boxwidth,2.5*\boxheight) 	-- (2.5*\boxwidth,-.5*\boxheight);

				\node 							(desc_temporal)	at (1.5*\boxwidth,-1*\boxheight-1*\boxoffset) {\textit{spatial}};
				\node[rotate=90, anchor=south] 	(desc_spatial) 	at (-.5*\boxwidth-1*\boxoffset,1.5*\boxheight) {\textit{temporal}};

				\draw[thick,<->] (-.5*\boxwidth-\boxoffset,.5*\boxheight) -- (-.5*\boxwidth-\boxoffset,2.5*\boxheight);
				\draw[thick,<->] (.5*\boxwidth,-.5*\boxheight-\boxoffset) -- (2.5*\boxwidth,-.5*\boxheight-\boxoffset);
			\end{tikzpicture}
		}
		\caption{Identified classes of hashtags within the Jodel platform according to the spatial and temporal spread metric.}
		\label{fig:hashtag_classes}
	\end{subfigure}
	\vspace*{-10px}
	\caption{The correlation between spatial and temporal spread.
		\emph{a)} describes the number of hashtags and their spread properties in the restricted dataset.
		\emph{b)} shows our derived classes of hashtags according to the spread metrics.
	}
	\label{fig:spread_combined}
	\vspace*{-20px}
\end{figure}

Having analyzed the spatial and temporal dimensions in isolation, we are now interested in how they correlate.
For example, hashtags that occur in one geographic area have a low spatial spread, but can be active over a short or longer timespan as indicated by the temporal spread.
Therefore, we focus on correlating the spatial and temporal spread and omit other metrics since they provide a similar picture.
Figure~\ref{fig:spreadvstemp} shows the spatial spread on the x-axis and the temporal spread on the y-axis of all hashtags having at least $30$ occurrences since 2016.
The hashtags can roughly be clustered into four groups as shown in Figure~\ref{fig:hashtag_classes}.
\emph{i)} A temporal spread of $100\,\text{days}$ and a spatial spread of $250\,\text{km}$ (long-lived and countrywide).
We would expect countrywide hashtags that are statements and also memes in this group, as both kinds are often spread out on the landscape and rather long-lived.
\emph{ii)} Located around a spatial spread of $250\,\text{km}$, but the temporal spread is only a few days (short-lived and global).
Hashtags in this group are, for example, about countrywide events.
Also, some memes that are short-lived could be in that group.
\emph{iii)} Spread around $0$ to $30\,\text{km}$ and temporal spread of $0$ to $70\,\text{days}$ (long-lived and local).
Here, we would expect hashtags about phenomena that are particularly local due to the community structure of Jodel.
\emph{iv)} Short-lived and local hashtags.
This group can involve for example local events.
We will base our content classification of hashtags in Section~\ref{sec:hashtag_classification} on these identified groups.

\afblock{Findings.}
The correlation of spatial and temporal spread clusters the hashtags into four groups, identified by long-lived vs.\ short-lived and countrywide/global vs.\ local spread.
That is, there are some long-lived and short-lived countrywide hashtags, while we also identify long- and short-lived local hashtag occurrences.

\subsection{Influence and Similarity of Cities}
We have seen that some hashtags occur rather locally, which is an essential aspect of the Jodel application.
We have also seen that many hashtags spread through many Jodel communities.
Therefore, we next want to examine how much communities influence each other in the sense of causing other cities to adopt a hashtag.
We are particularly interested in which cities source and popularize trends before others adopt them.

\afblock{Spatial impact.}
To get insights of on cities' impact on another, we use the spatial impact metric from \cite{spatiotemp}.
The hashtag specific spatial impact $I_{A\rightarrow B}^h$ of two cities $A$ and $B$ and a hashtag $h$ is defined as a score in the range $[-1,1]$.
A score of $1$ means that either all occurrences of that hashtag in city $A$ happened before all occurrences in $B$, or that there are no occurrences of that hashtag in $B$ at all.
The same applies in the reverse case scoring $-1$.
Values around $0$ indicate that both cities adopted the hashtag roughly at the same time.
In short, this measure describes which city adopted a hashtag earlier, and therefore \emph{may} have influenced the other city.
The spatial impact $I_{A\rightarrow B}$ is then defined as the average hashtag's spatial impact {\em for all hashtags} that occur in at least one of the cities.

\begin{figure}
	\centering
	\begin{subfigure}[c]{0.32\linewidth}
		\includegraphics[width=1\linewidth]{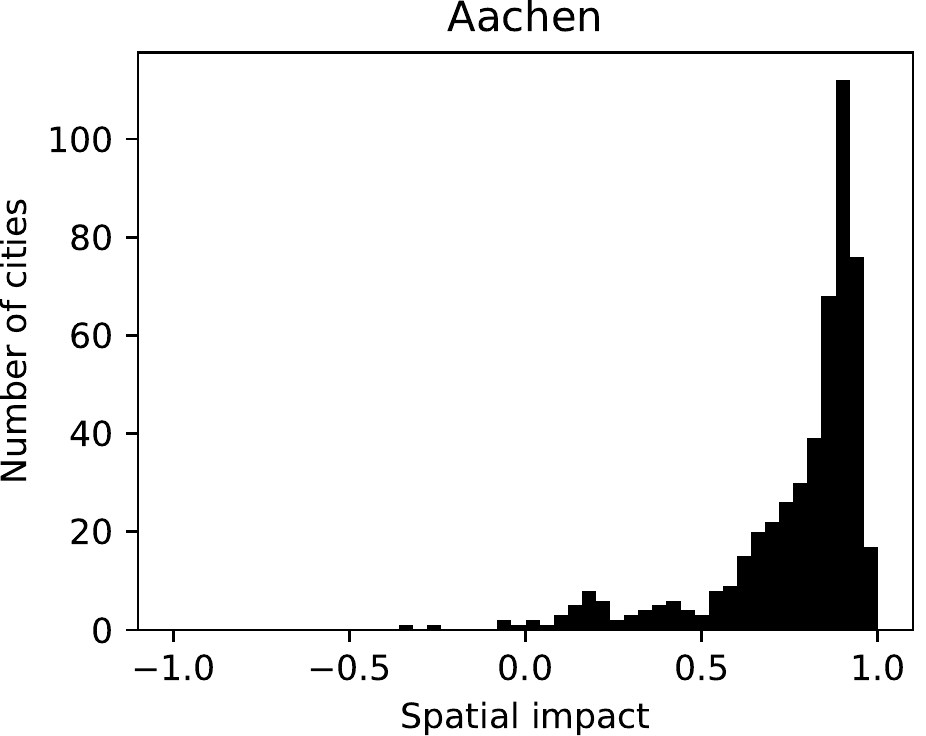}
		\subcaption{Aachen}
	\end{subfigure}
		\begin{subfigure}[c]{0.32\linewidth}
		\includegraphics[width=1\linewidth]{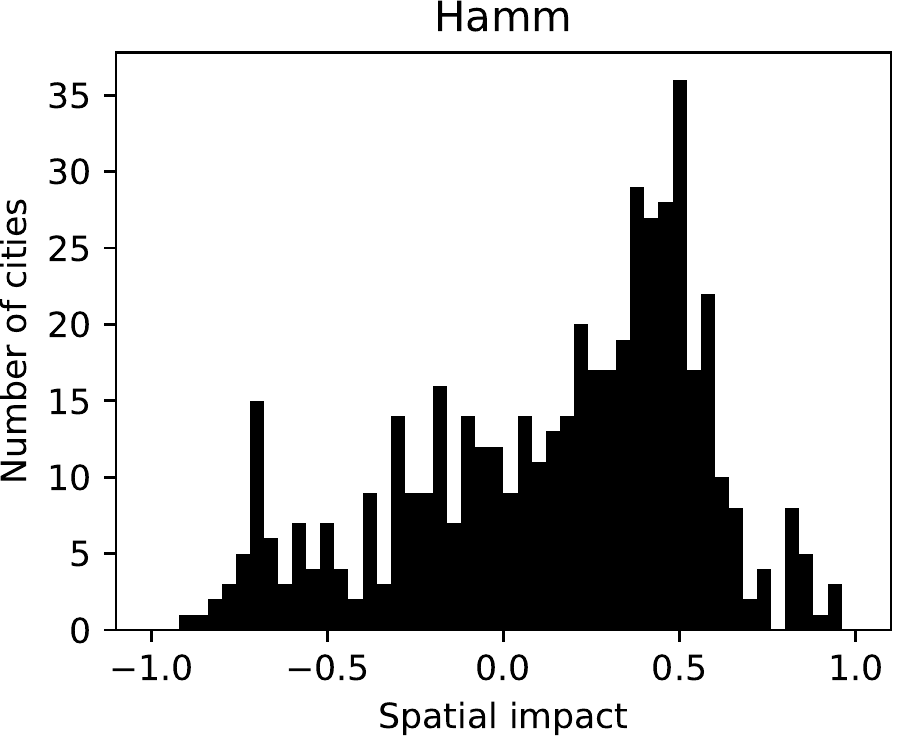}
	\subcaption{Hamm}
	\end{subfigure}
		\begin{subfigure}[c]{0.32\linewidth}
		\includegraphics[width=1\linewidth]{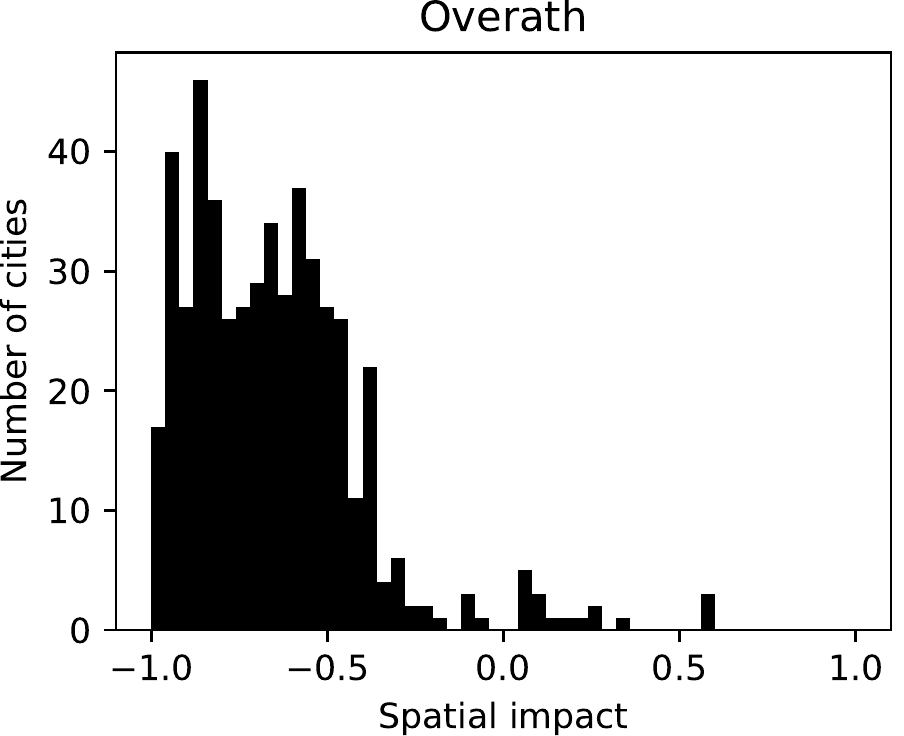}
	\subcaption{Overath}
	\end{subfigure}
    \vspace*{-10px}
	\caption{Histograms of spatial impact from Aachen, Hamm, and Overath to the top 500 locations in the complete dataset.
		Aachen heavily influences most other cities, Overath is mostly influenced by other cities, and Hamm is both influenced by several cities and influencing other cities.}
	\label{fig:si}
    \vspace*{-.5cm}
\end{figure}

As an example, we compare the cities Aachen, Hamm, and Overath with the 500 most popular cities.
For each of the three cities, we show the spatial impact on every of the 500 most popular cities as a histogram in Figure~\ref{fig:si}.
We chose Aachen as the birthplace of the Jodel network with a large technical university and 250\,k inhabitants, Hamm as a medium-sized city without university and 180\,k inhabitants, and Overath as a smaller city with 27\,k inhabitants.
The histograms x-axis denotes the spatial impact, while the y-axis covers the number of other cities in comparison.
From the given examples, we observe that Aachen is the most influencing city within this comparison (and also on the whole platform Jodel--\emph{not shown}), with most of its scores being between $0.5$ and $1$.
Hamm is both influenced by cities as well as influencing other cities, whereas Overath is heavily influenced by most other cities (probably also due to a low population and therefore fewer users).
By also qualitatively looking into other cities spatial impact histogram, we can only conclude that cities with a higher population impact cities with a lower population.
This finding that large cities influence smaller ones is in line with observations on Twitter~\cite{spatiotemp}.

We remark that the spatial impact metric does not normalize by community size and thus comparing communities of unequal size can provide an advantage in this metric to the larger city.
Even if the hashtags in the big city never spread to any other city, it would still impact a small city using this measure.
Nevertheless, this still supports the findings also shown for Twitter that larger cities usually have a higher impact.

\afblock{Hashtag similarity.}
We previously have seen that cities impact each other.
To understand the communities hashtags better in comparison, we use the \textit{hashtag similarity} \cite{spatiotemp} measure of two locations $A$ and $B$ as
$\text{sim}(A, B) = {\vert H_A^{50} \cap H_B^{50} \vert}/{50}$,
where $H_L^{50}$ defines the 50 most popular hashtags in location $L$.

\begin{figure}
	\centering
	\begin{subfigure}[c]{0.32\linewidth}
		\includegraphics[width=1\linewidth]{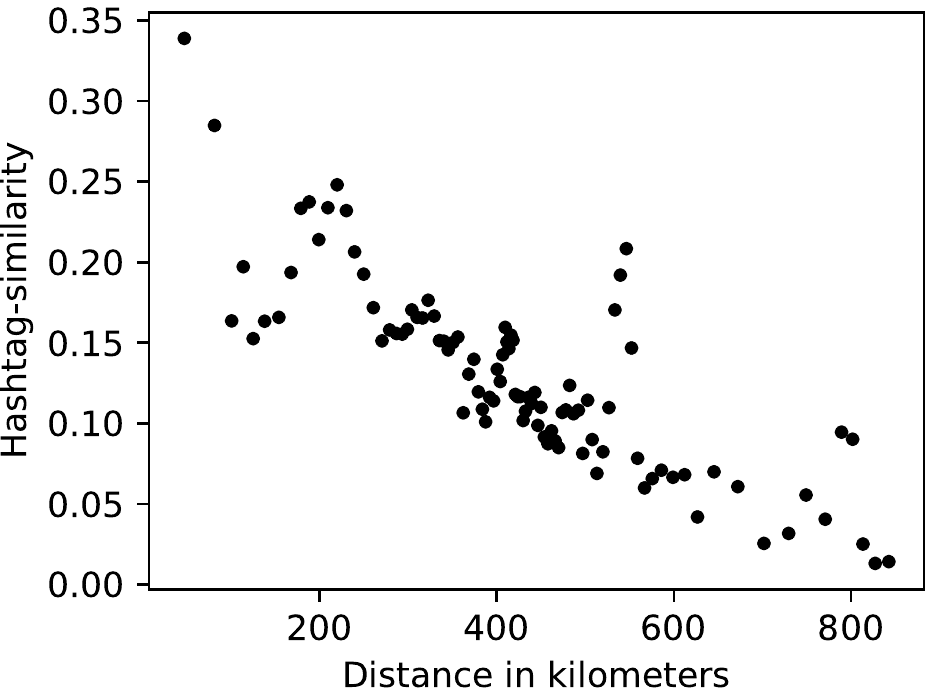}
		\subcaption{Aachen}
	\end{subfigure}
	\begin{subfigure}[c]{0.32\linewidth}
		\includegraphics[width=1\linewidth]{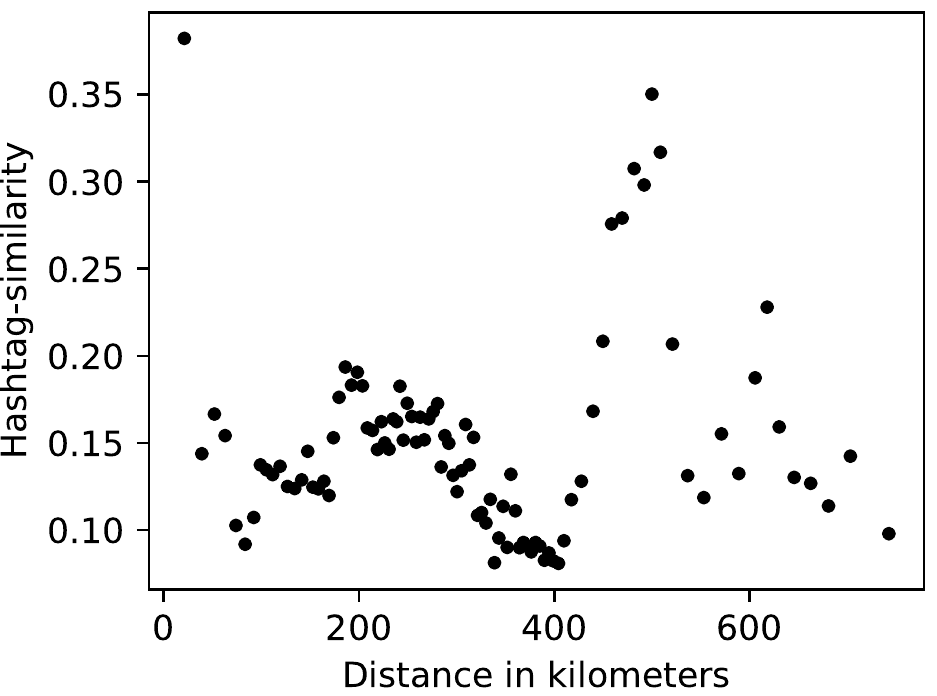}
		\subcaption{Munich}
	\end{subfigure}
	\begin{subfigure}[c]{0.32\linewidth}
		\includegraphics[width=1\linewidth]{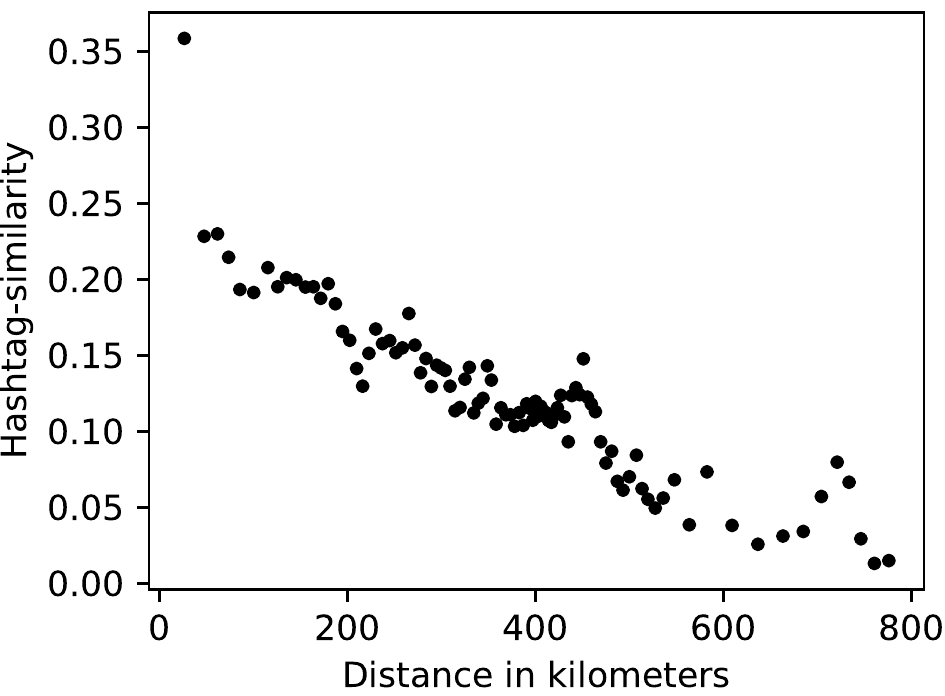}
		\subcaption{Overath}
	\end{subfigure}
    \vspace*{-10px}
	\caption{The hashtag similarity of Aachen, Munich, and Overath to cities in a certain distance.
		Cities closer to each other tend to share more hashtags.
		However, big cities are similar to each other no matter the distance.
		Averages of groups of 100 locations.}
	\label{fig:sim}
    \vspace*{-10px}
\end{figure}

For each location, we calculated the hashtag similarity to all others.
Figure~\ref{fig:sim} shows the results for Aachen, Munich, and Overath in averages for groups of 100 locations. %
While the x-axis describes the distance to other cities, the y-axis denotes the similarity score.
For Aachen and Overath, we observe that closer locations are on average more similar than locations farther away.
However, there are several peaks of which the biggest ones represents Berlin\footnote{Within our dataset, Berlin is split into districts and therefore present multiple times.}.
It seems apparent that big cities are connected to each other and share hashtags no matter the distance, which is supported by the example of Munich.
Yet, small cities like Overath are less affected.
\cite{butterflies} showed similar results for Twitter: W.r.t hashtags, big cities are more similar to each other than to closer, smaller cities.

We verified that this also applies for Jodel considering all hashtags of both cities.
The relation we see for Overath of closer cities having more hashtags in common has likewise been shown for Twitter~\cite{spatiotemp}.
Our hypothesis is that on Jodel, hashtags \emph{travel} long distances between big cities and then spread across smaller cities within the local neighborhood.

\afblock{Findings.}
While the hashtag similarity metric does not directly reflect individual user's contribution to hashtag spreading, it still provides insights into the dis-/similar hashtag usage of communities.
Big cities share more popular hashtags and are therefore generally more similar to each other, whereas smaller cities gradually share their most popular hashtags with their local neighborhood.
In combination with the spatial influence, this supports our conclusion that hashtags likely spread via the bigger cities into such local neighborhoods.

\section{Hashtag Classification}
\label{sec:hashtag_classification}
Within our analysis of hashtags, we have observed that the hashtags can be clustered into different groups (cf. Figure~\ref{fig:spreadvstemp} \& \ref{fig:hashtag_classes}).
We know from literature that there are corresponding types of hashtags on \eg Twitter.
That is, \cite{spatiotemp} distinguishes between local interest hashtags, regional and event-driven hashtags, and other worldwide memes.
We were wondering if and in which way Jodel's locality actually catalyzes other---very local---or prohibits global hashtags.
For answering this questions, we create a statistical classifier for determining the hashtag type in three steps: \emph{i)} defining suitable hashtag classes in line with our observations so far, \emph{ii)} manual hashtag classification for providing an answer on a content level, and \emph{iii)} training and validation of statistical models.

\subsection{Hashtag Content Categories}
Leveraging hints from Section~\ref{sec:spatial_vs_temporal}, manual inspection and expert domain knowledge, we first iteratively defined and verified four different meme classes as follows:
\begin{itemize}[noitemsep,topsep=5pt,leftmargin=9pt]
	\item \textbf{Local events:} Often trends originating from a single post (\eg a funny story) that gained attention in the local community.
	It is typically very local and short-lived.
	\item \textbf{Local phenomena:}	Trend usually related to local persons or buildings.
	It is typically very local and long-lived.
	\item \textbf{Events:} Short-lived or recurring trend usually related to a real-world happening of larger interest.
	\item \textbf{Other memes:} Memes not included in Jodelstories or Local phenomena.
\end{itemize}

We labeled the most 450 popular hashtags that had their first occurrence after 1st January 2016 to filter out most of the generic statements.
Besides, this makes the classes more balanced, as local trends are much more prominent in this restricted dataset.
Due to missing context information or non-fitting classes, we could not classify 49 hashtags. %
The majority (64\,\%) of the remaining 401 hashtags were labeled \emph{other meme}, whereas \emph{local phenomenon} (82) represents the second biggest class, Events (35) and Local Event (29) being relatively equal in size.

Having learned that we indeed find trends in terms of hashtags that w.r.t our previous metrics and the manual classification reflect the locality of the Jodel application, we next try to establish the classification methods for them.
Thus, we define features that we will use including the presented and analyzed metrics plus some additional temporal and text-based ones in the following section.

\subsection{Features}
Our aim is to create a statistical classifier for determining the hashtag type.
For our classification approach, we used the features listed in Table~\ref{table:features}.
This list includes all spatial and temporal metrics that have been discussed before.
Besides simple features like hashtag and comment counts, we further added temporal metrics of \textit{peak increase} being defined as the number of posts in seven days prior to the peak divided by the number of posts on the peak day---and \textit{peak decline} alike, but after the peak.
These features, therefore, describe how suddenly a trend occurred and disappeared.

\begin{table}
	\caption{
        \vspace*{-10px}
        The features used for the classification.}
	\label{table:features}
	\footnotesize
	\begin{tabular}{|l|p{6.2cm}|}
		\hline
		\textbf{Feature}          & \textbf{Definition} \\ \hline\hline
		Focus            & The focus of the hashtag. \\ \hline
		Entropy          & The entropy of the hashtag. \\ \hline
		Spread           & The spread of the hashtag. \\ \hline
		Local variation  & The local variation of the hashtag. A measure for the regularity of the hashtag's usage. \\ \hline
		Hashtags         & Average number of hashtags per Jodel. \\ \hline
		Comments         & Average number of comments per Jodel. \\ \hline
		Exclamations     & Fraction of Jodels that contain an exclamation mark. \\ \hline
		Questions        & Fraction of Jodels that contain a question mark. \\ \hline
		Temporal focus   & The amount of Jodels posted on the peak day of the hashtag divided by the total number of uses. \\ \hline
		Temporal entropy & Similar to spatial entropy where different days are considered. Gives a number for the ``randomness'' of the distribution. \\ \hline
		Temporal spread  & Similar to spatial spread of the avg distance [days] from the weighted midpoint of all occurrences of the hashtag. \\ \hline
		Peak increase    & Compares post volume of seven days before the peak with the height of the peak. Is a measure for how ``sudden'' the peak occurred. A low value indicates a sudden increase in popularity.  \\ \hline
		Peak decline     & Seven days after the peak divided by the height of the peak. Describes how fast interest declined after the peak day. A low value means the interest disappeared suddenly. \\ \hline
		User diversity   & Number of unique users of the hashtag divided by its total use. \\ \hline
	\end{tabular}
\end{table}

\subsection{Classifiers and Results}
\afblock{Classifiers.}
We have applied different statistical methods to our classification problem: k-nearest neighbors, Classification \& Regression Trees, Naive Bayes, Logistic Regressen, LDA and ZeroR as a baseline. %
We used 10-fold cross-validation on our manually classified hashtag dataset to verify the results of each classifier. %
All classifiers outperform the baseline ZeroR-classifier.
While all approaches perform well (detailed results omitted), LDA resulted in a good compromise of the smallest average $\pm$ standard deviation.
Therefore, we only present the results of the LDA classifier in Table~\ref{table:results2}.
We observe that {\em events} have the lowest precision value with $0.66$.
However, this is still a good result as less than $10$\% of the hashtags are events.
The other results are good as well, especially the local phenomena and memes with high F1 scores.

In this classification, both the spatial and the temporal features provided most benefit as removing them caused in both cases a considerable drop in accuracy of at least $0.1$, whereas user diversity had only a very minor influence. %

\afblock{Findings.}
We have shown that we can predict the class of a hashtag by using its spatial and temporal properties.
In conclusion, this confirms our theory that the Jodel platform actually has specific local short-lived and long-lived hashtags that differ to countrywide generic memes and events.
While we may extend the classification scheme with more features and could apply advanced machine learning techniques, such as neural networks, this is a first step towards automatically classifying certain countrywide/gloabl and in opposition local trends on Jodel---either being short- or long-lived according to our defined classes.

\begin{table}
    \vspace*{-.5cm}
		\caption{Precision, recall and f1-score using the LDA classifier.
			Averages of 10 runs with different dataset-splits.
            \vspace*{-10px}
            }
		\label{table:results2}
		\footnotesize
		\begin{tabular}{|l|c|c|c|}
			\hline
							& \textbf{Precision} & \textbf{Recall} & \textbf{F1-Score}  \\ \hline\hline
			Event			& 0.66      & 0.80   & 0.70 \\ \hline
			Local event		& 0.79      & 0.72   & 0.74 \\ \hline
			Local phenomena	& 0.87      & 0.95   & 0.91 \\ \hline
			Other memes		& 0.97      & 0.93   & 0.94 \\ \hline
		\end{tabular}
    \vspace*{-.5cm}
\end{table}

\vspace*{-5px}
\section{Conclusions}
\label{sec:future_work_and_conclusion}

Within this paper, we study the hashtag propagation through the lens of a platform operator by having the unique opportunity to analyze data from Germany (2014 to 2017) provided by Jodel.
With this longitudinal data set, we studied the key design pattern of being location-based and its influence on hashtag usage and spreading in comparison to the global counterpart Twitter.
We applied established metrics designed to capture the spatial focus and spread of Twitter hashtags [2] to Jodel and extend them with a temporal dimension covering the diffusion of hashtags in time.
While we find significant qualitative differences to Twitter of hashtags generally being less focused on Jodel and thus having a higher entropy, the spatial spread also deviates from Twitter.
Yet, we find evidence for local hashtags that are a potential result of Jodel's design.

Further, we identify similarities in hashtag usage between nearby and larger cities and present case studies of their spacial impact supporting this finding.
By correlating spatial and temporal metrics, we identify four different hashtag classes distinguished by their spatial and temporal extent.
Informed by manual labeling of 450 most frequently used hashtags, we created an automatic classification scheme using machine learning models with great success.

While we focused on the empirical birds-eye view on the hashtag usage, it will be interesting trying to apply epidemic modeling approaches.
Further, individual user behavior and possible groups w.r.t their spreading influence will provide deeper insights---especially in the sense of Jodel's design choice of being location-based.

\afblock{Acknowledgement}\\
This work has been funded by the Execellence Initiative of the German federal and state governments.

\vspace*{-5px}

\bibliographystyle{ACM-Reference-Format}
\balance
\bibliography{paper}


\begin{thebibliography}{20}


\ifx \showCODEN    \undefined \def \showCODEN     #1{\unskip}     \fi
\ifx \showDOI      \undefined \def \showDOI       #1{#1}\fi
\ifx \showISBNx    \undefined \def \showISBNx     #1{\unskip}     \fi
\ifx \showISBNxiii \undefined \def \showISBNxiii  #1{\unskip}     \fi
\ifx \showISSN     \undefined \def \showISSN      #1{\unskip}     \fi
\ifx \showLCCN     \undefined \def \showLCCN      #1{\unskip}     \fi
\ifx \shownote     \undefined \def \shownote      #1{#1}          \fi
\ifx \showarticletitle \undefined \def \showarticletitle #1{#1}   \fi
\ifx \showURL      \undefined \def \showURL       {\relax}        \fi
\providecommand\bibfield[2]{#2}
\providecommand\bibinfo[2]{#2}
\providecommand\natexlab[1]{#1}
\providecommand\showeprint[2][]{arXiv:#2}

\bibitem[\protect\citeauthoryear{Becker, Naaman, and Gravano}{Becker
  et~al\mbox{.}}{2011}]%
        {class_2}
\bibfield{author}{\bibinfo{person}{Hila Becker}, \bibinfo{person}{Mor Naaman},
  {and} \bibinfo{person}{Luis Gravano}.} \bibinfo{year}{2011}\natexlab{}.
\newblock \showarticletitle{{Beyond Trending Topics: Real-World Event
  Identification on Twitter}} \emph{(\bibinfo{series}{ICWSM})}.
\newblock


\bibitem[\protect\citeauthoryear{Cannarella and Spechler}{Cannarella and
  Spechler}{2014}]%
        {epidemics_1}
\bibfield{author}{\bibinfo{person}{John Cannarella} {and}
  \bibinfo{person}{Joshua~A. Spechler}.} \bibinfo{year}{2014}\natexlab{}.
\newblock \showarticletitle{Epidemiological modeling of online social network
  dynamics} \emph{(\bibinfo{series}{CoRR})}.
\newblock


\bibitem[\protect\citeauthoryear{Chandra, Khan, and Muhaya}{Chandra
  et~al\mbox{.}}{2011}]%
        {user_estimation}
\bibfield{author}{\bibinfo{person}{S. Chandra}, \bibinfo{person}{L. Khan},
  {and} \bibinfo{person}{F.~B. Muhaya}.} \bibinfo{year}{2011}\natexlab{}.
\newblock \showarticletitle{{Estimating Twitter User Location Using Social
  Interactions--A Content Based Approach}}
  \emph{(\bibinfo{series}{SocialCom/PASSAT})}.
\newblock


\bibitem[\protect\citeauthoryear{Dow, Adamic, and Friggeri}{Dow
  et~al\mbox{.}}{2013}]%
        {facebook}
\bibfield{author}{\bibinfo{person}{P.~Alex Dow}, \bibinfo{person}{Lada~A.
  Adamic}, {and} \bibinfo{person}{Adrien Friggeri}.}
  \bibinfo{year}{2013}\natexlab{}.
\newblock \showarticletitle{The Anatomy of Large Facebook Cascades}
  \emph{(\bibinfo{series}{ICWSM})}.
\newblock


\bibitem[\protect\citeauthoryear{Ferrara, Varol, Menczer, and Flammini}{Ferrara
  et~al\mbox{.}}{2013}]%
        {butterflies}
\bibfield{author}{\bibinfo{person}{Emilio Ferrara}, \bibinfo{person}{Onur
  Varol}, \bibinfo{person}{Filippo Menczer}, {and} \bibinfo{person}{Alessandro
  Flammini}.} \bibinfo{year}{2013}\natexlab{}.
\newblock \showarticletitle{{Traveling Trends: Social Butterflies or Frequent
  Fliers?}} \emph{(\bibinfo{series}{COSN})}.
\newblock


\bibitem[\protect\citeauthoryear{Kamath, Caverlee, Lee, and et~al.}{Kamath
  et~al\mbox{.}}{2013}]%
        {spatiotemp}
\bibfield{author}{\bibinfo{person}{Krishna~Y Kamath}, \bibinfo{person}{James
  Caverlee}, \bibinfo{person}{Kyumin Lee}, {and} \bibinfo{person}{et al.}}
  \bibinfo{year}{2013}\natexlab{}.
\newblock \showarticletitle{{Spatio-Temporal Dynamics of Online Memes: A Study
  of Geo-Tagged Tweets}} \emph{(\bibinfo{series}{WWW})}.
\newblock


\bibitem[\protect\citeauthoryear{Kotsakos, Sakkos, Katakis, and
  Gunopulos}{Kotsakos et~al\mbox{.}}{2014}]%
        {class_3}
\bibfield{author}{\bibinfo{person}{Dimitrios Kotsakos}, \bibinfo{person}{Panos
  Sakkos}, \bibinfo{person}{Ioannis Katakis}, {and} \bibinfo{person}{Dimitrios
  Gunopulos}.} \bibinfo{year}{2014}\natexlab{}.
\newblock \showarticletitle{{\#Tag: Meme or Event?}}
  \emph{(\bibinfo{series}{ASONAM})}.
\newblock


\bibitem[\protect\citeauthoryear{Leskovec, Backstrom, and Kleinberg}{Leskovec
  et~al\mbox{.}}{2009}]%
        {blogosphere}
\bibfield{author}{\bibinfo{person}{Jure Leskovec}, \bibinfo{person}{Lars
  Backstrom}, {and} \bibinfo{person}{Jon Kleinberg}.}
  \bibinfo{year}{2009}\natexlab{}.
\newblock \showarticletitle{{Meme-tracking and the Dynamics of the News Cycle}}
  \emph{(\bibinfo{series}{KDD})}.
\newblock


\bibitem[\protect\citeauthoryear{Li, Lei, Khadiwala, and Chang}{Li
  et~al\mbox{.}}{2012}]%
        {detection_1}
\bibfield{author}{\bibinfo{person}{R. Li}, \bibinfo{person}{K.~H. Lei},
  \bibinfo{person}{R. Khadiwala}, {and} \bibinfo{person}{K.~C. Chang}.}
  \bibinfo{year}{2012}\natexlab{}.
\newblock \showarticletitle{{TEDAS: A Twitter-based Event Detection and
  Analysis System}} \emph{(\bibinfo{series}{ICDE})}.
\newblock


\bibitem[\protect\citeauthoryear{Mahler}{Mahler}{2015}]%
        {yikyakNyTimes}
\bibfield{author}{\bibinfo{person}{Jonathan Mahler}.}
  \bibinfo{year}{2015}\natexlab{}.
\newblock \showarticletitle{{Who Spewed That Abuse? Yik Yak Isn’t Telling}}.
\newblock
  \bibinfo{howpublished}{\url{http://www.nytimes.com/images/2015/03/09/nytfrontpage/scan.pdf}}.
\newblock  (\bibinfo{year}{2015}).
\newblock


\bibitem[\protect\citeauthoryear{Matsubara, Sakurai, and et~al.}{Matsubara
  et~al\mbox{.}}{2017}]%
        {model_1}
\bibfield{author}{\bibinfo{person}{Yasuko Matsubara}, \bibinfo{person}{Yasushi
  Sakurai}, {and} \bibinfo{person}{et al.}} \bibinfo{year}{2017}\natexlab{}.
\newblock \showarticletitle{{Nonlinear Dynamics of Information Diffusion in
  Social Networks}}.
\newblock \bibinfo{journal}{\emph{ACM Trans. Web}}.
\newblock


\bibitem[\protect\citeauthoryear{Romero, Meeder, and Kleinberg}{Romero
  et~al\mbox{.}}{2011}]%
        {topics}
\bibfield{author}{\bibinfo{person}{Daniel~M. Romero}, \bibinfo{person}{Brendan
  Meeder}, {and} \bibinfo{person}{Jon Kleinberg}.}
  \bibinfo{year}{2011}\natexlab{}.
\newblock \showarticletitle{{Differences in the Mechanics of Information
  Diffusion Across Topics: Idioms, Political Hashtags, and Complex Contagion on
  Twitter}} \emph{(\bibinfo{series}{WWW})}.
\newblock


\bibitem[\protect\citeauthoryear{Sakaki, Okazaki, and Matsuo}{Sakaki
  et~al\mbox{.}}{2010}]%
        {earthquake}
\bibfield{author}{\bibinfo{person}{Takeshi Sakaki}, \bibinfo{person}{Makoto
  Okazaki}, {and} \bibinfo{person}{Yutaka Matsuo}.}
  \bibinfo{year}{2010}\natexlab{}.
\newblock \showarticletitle{{Earthquake Shakes Twitter Users: Real-time Event
  Detection by Social Sensors}} \emph{(\bibinfo{series}{WWW})}.
\newblock


\bibitem[\protect\citeauthoryear{Sanlı and Lambiotte}{Sanlı and
  Lambiotte}{2015}]%
        {local_variation}
\bibfield{author}{\bibinfo{person}{Ceyda Sanlı} {and} \bibinfo{person}{Renaud
  Lambiotte}.} \bibinfo{year}{2015}\natexlab{}.
\newblock \showarticletitle{Local Variation of Hashtag Spike Trains and
  Popularity in Twitter}.
\newblock \bibinfo{journal}{\emph{PLOS ONE}} (\bibinfo{year}{2015}).
\newblock


\bibitem[\protect\citeauthoryear{Walther and Kaisser}{Walther and
  Kaisser}{2013}]%
        {detection_3}
\bibfield{author}{\bibinfo{person}{Maximilian Walther} {and}
  \bibinfo{person}{Michael Kaisser}.} \bibinfo{year}{2013}\natexlab{}.
\newblock \showarticletitle{{Geo-spatial Event Detection in the Twitter
  Stream}} \emph{(\bibinfo{series}{Advances in Information Retrieval})}.
\newblock


\bibitem[\protect\citeauthoryear{Weng and Lee}{Weng and Lee}{2011}]%
        {detection_2}
\bibfield{author}{\bibinfo{person}{Jianshu Weng} {and} \bibinfo{person}{Bu-Sung
  Lee}.} \bibinfo{year}{2011}\natexlab{}.
\newblock \showarticletitle{{Event Detection in Twitter}}
  \emph{(\bibinfo{series}{ICWSM})}.
\newblock


\bibitem[\protect\citeauthoryear{Woo and et~al.}{Woo and et~al.}{2016}]%
        {epidemics_3}
\bibfield{author}{\bibinfo{person}{Jiyoung Woo} {and} \bibinfo{person}{et al.}}
  \bibinfo{year}{2016}\natexlab{}.
\newblock \showarticletitle{{Epidemic model for inform. diffusion in web
  forums: experiments in marketing exchange and political dialog}}.
\newblock \bibinfo{journal}{\emph{SpringerPlus}} (\bibinfo{year}{2016}).
\newblock


\bibitem[\protect\citeauthoryear{Xu and et~al.}{Xu and et~al.}{2016}]%
        {youtube}
\bibfield{author}{\bibinfo{person}{Weiai~Wayne Xu} {and} \bibinfo{person}{et
  al.}} \bibinfo{year}{2016}\natexlab{}.
\newblock \showarticletitle{{Networked Cultural Diffusion and Creation on
  YouTube: An Analysis of YouTube Memes}}.
\newblock \bibinfo{journal}{\emph{J. of Broadc. \& Electr. Media}}
  (\bibinfo{year}{2016}).
\newblock


\bibitem[\protect\citeauthoryear{Yan, Wu, and et~al.}{Yan
  et~al\mbox{.}}{2013}]%
        {epidemics_2}
\bibfield{author}{\bibinfo{person}{Qiang Yan}, \bibinfo{person}{Lianren Wu},
  {and} \bibinfo{person}{et al.}} \bibinfo{year}{2013}\natexlab{}.
\newblock \showarticletitle{Information Propagation in Online Social Network
  Based on Human Dynamics} \emph{(\bibinfo{series}{Abstract and Applied
  Analysis})}.
\newblock


\bibitem[\protect\citeauthoryear{Zannettou and et~al.}{Zannettou and
  et~al.}{2017}]%
        {zannettou2017web}
\bibfield{author}{\bibinfo{person}{Savvas Zannettou} {and} \bibinfo{person}{et
  al.}} \bibinfo{year}{2017}\natexlab{}.
\newblock \showarticletitle{{The web centipede: underst. how web communit.
  infl. each other through the lens of mainstream and alt. news sources}}
  \emph{(\bibinfo{series}{IMC})}.
\newblock


\end{thebibliography}

\end{document}